\documentclass[preprint,showpacs,aps,draft]{revtex4}
\usepackage{graphicx}
\usepackage{dcolumn}
\usepackage{longtable}
\begin{document}
\title{{\Large  Structure of  A=138 isobars above  the  $^{132}Sn$
core  and  the N-N interaction in the neutron-rich environment}}

\author{S. Sarkar}
\affiliation{Department   of   Physics,  Bengal  Engineering  and
Science University, Shibpur, Howrah - 711103, INDIA.}

\author{M. Saha Sarkar
\thanks{e-mail: maitrayee.sahasarkar@saha.ac.in}}
\affiliation{Nuclear  and Atomic Physics Division, Saha Institute
of Nuclear Physics, Kolkata 700064, INDIA }

\date{\today}

\begin{abstract}
Large  basis  untruncated shell model calculations have been done
for the A=138 neutron  -rich  nuclei  in  the  $\pi(gdsh)  \oplus
\nu(hfpi)$  valence  space  above  the $^{132}Sn$ core. Two (1+2)
-body nuclear Hamiltonians, viz.,  realistic  CWG  and  empirical
SMPN  in this model space have been used. Calculated ground state
binding  energies,  level   spectra   and   other   spectroscopic
properties  have  been  compared  with the available experimental
data. Importance of untruncated shell model calculations in  this
model  space  has  been  pointed out. Shell model results for the
very  neutron  rich  Sn   isotope   ($^{138}Sn$,   N/Z=1.76)   of
astrophysical  interest  for  which  no spectroscopic information
except  $\beta  $  -decay  half  life  is  available,  have  been
presented.  Shell  structure and evolution of collectivity in the
even-even A=138 isobars  have  been  studied  as  a  function  of
valence neutron and /or proton numbers. Calculations done for the
first  time, reproduce remarkably well the collective vibrational
states in $^{138}Te$ and $^{138}Xe$. Comparison of  some  of  the
important  two-body matrix elements of the empirical SMPN, CW5082
and the realistic CWG interactions has been  done.  These  matrix
elements  are  important  for  ground  state binding energies and
low-lying spectra of nuclei in this region. Consideration of  the
predictability  of the two interactions seems to suggest that, in
order to incorporate the special features of the N-N  interaction
in  such exotic n-rich environment above the $^{132}Sn$ core, the
use of local spectroscopic information from the region  might  be
essential.

\end{abstract}

\maketitle
\section{Introduction}
Recent   experimental   and  theoretical  investigations  in  the
close-to-neutron-drip  line  region  above   the   doubly   magic
$^{132}Sn$   nucleus   have   revealed   many  intriguing  issues
concerning some newer aspects of nuclear structure and  empirical
/ realistic N-N interaction in such exotic environments. Decrease
in  the  spin-orbit  interaction  at  N-Z=32 above the Z=50, N=82
double shell closure  \cite{sch:1},  anomalously  low  excitation
energy  (282  keV)  of  the first 5/2$^+$ state in the $^{135}Sb$
compared to its position in $^{133}Sb$  (962  keV)  \cite{she:1},
anomalously  low  B($E2,  0^+_{gs}  \rightarrow  2^+_1$) value in
$^{136}Te$  \cite{rad:1},   the   question   of   appearance   of
deformation  and  the locus of transition from single particle to
collective behaviour, as well as the evolution of collectivity in
these neutron-rich nuclei above  $^{132}Sn$  core  and  the  role
therein  of  the number of valence neutrons and protons, are some
of the structure issues being studied recently. The  question  of
reduced  neutron-neutron pairing in the exotic n-rich environment
\cite{ritu:1} above N=82 \cite{tera:1}, and modification  of  the
empirical  /  effective  two-body  matrix elements (tbmes) in the
valence space ($\pi gdsh$, $\nu hfpi$) above the $^{132}Sn$  core
for  shell  model  applications,  are  also  of current interest.
Moreover nuclei with 50$\leq$Z$\leq$56 and 82$\leq$N$\leq$88  lie
on or close to the path of the astrophysical r-process flow. Thus
the   knowledge   of  the  structure,  particularly  the  binding
energies, low - lying excited states and $\beta$-decay  rates  of
these  nuclei  at  finite temperatures, are important ingredients
for investigation in the nucleosynthesis scenario. However,  many
of  these  important nuclei are yet to be studied experimentally.
For  example,  the  even-N  neutron-rich  Sn  isotopes  are   the
classical  "waiting  point"  nuclei  in  the  A=130  solar system
abundance peak under typical r-process conditions.  With  respect
to  the  r-process,  $^{136}Sn$  is  an  important waiting -point
nucleus for moderate neutron densities  to  drive  the  r-process
flow  beyond  A$\simeq$130  peak  \cite{aw:1}. Some spectroscopic
information, such as binding energy and low -lying spectrum exist
only for the $^{134}Sn$. Recently, the $\beta$-decay  half  lives
via delayed neutron emission have been measured for $^{135}Sn$ to
$^{138}Sn$   nuclei   \cite{aw:1}.  Their  production  rates  and
lifetimes  are  small.  This  provides  severe   limitations   in
acquiring  spectroscopic information on them. Therefore, reliable
theoretical calculations and their comparisons with the  hitherto
available  experimental  data  on  the  binding  energies,  level
spectra, electromagnetic properties and $\beta$ -decay rates  are
not only important for extending our knowledge of N-N interaction
and  nuclear structure in the exotic n-rich environment, but they
also provide very useful ingredients for reviewing  the  problems
related to the nuclear astrophysics in general.

Theoretical  studies in the framework of nuclear shell model (SM)
have                          been                           done
\cite{sar:1,dae2002,dean:1,epja,sar:2,sar:3,cor:1,bro:1,136i:1,138i:1,138cs:1,kar:1}
recently  for  these  nuclei  above  the  $^{132}Sn$  core in the
valence space consisting of  $\pi(1g_{7/2},  2d_{5/2},  2d_{3/2},
3s_{1/2},  1h_{11/2})$  and  $\nu(1h_{9/2},  2f_{7/2},  2f_{5/2},
3p_{3/2}, 3p_{1/2}, 1i_{13/2})$ orbitals. These  SM  calculations
using  realistic  \cite{dean:1,cor:1,bro:1,kar:1}  and  empirical
\cite{sar:1,dae2002,epja,sar:2,sar:3,136i:1,138i:1,138cs:1}
interactions  in this valence space yield remarkably good results
for some of the isotopes of Sn, Sb, Te, I, Xe, Cs. The  empirical
SMPN  \cite{dae2002,epja} interaction, applied for all the nuclei
in the range 134 $\leq$ A $\leq$ 138 and $50\leq$Z$\leq  56$  are
found   to   be   highly  successful  \cite{epja}  in  predicting
particularly the ground state  spin-parities,  binding  energies,
level  spectra and electromagnetic transition probabilities. In a
recent  work  \cite{epja}  we  have  identified   that   a   mild
collectivity  appears  in  the $^{137}I,Te$ \cite{137i:1,137te:1}
nuclei  with  N  =  84  and  85,  respectively.  The  spectra  of
$^{138}Te$    \cite{138te:1},   $^{138}Xe$   \cite{138xe:1}   and
$^{139}I$  \cite{139i:1,nndc}  also   showed   good   vibrational
characteristics. Because of the computational limitations we then
studied  the  evolution  of  collectivity in the odd-A nuclei for
N=84-87 with Particle-Rotor Model \cite{khe:1} and indicated  the
role played by the valence protons and neutrons in them. The role
of the valence protons and neutrons have been also studied in the
context   of   the   structure   of   $^{138}I$   and  $^{138}Cs$
\cite{138i:1,138cs:1}. It has been found that the single particle
behaviour persists for nuclei with neutrons only in  the  valence
space  up  to  at  least  N=87.  The  calculated  results for the
spectroscopic information on  $^{136,137}$Sn,  Sb  isotopes  have
been  reported already in Ref.\cite{epja,sar:3}. The present work
is motivated to discuss the results of  untruncated  shell  model
calculations  over the $\pi(gdsh) \oplus \nu(hfpi)$ valence space
on the spectroscopic properties of the  A=138  even-even  isobars
and  to study the evolution of collectivity in these nuclei using
the code OXBASH \cite{oxb:1}. It is expected to reveal  the  role
of  valence nucleon species on the evolution of collectivity. The
other motivation is to compare the  results  of  SM  calculations
with  SMPN  interaction  \cite{epja}  with those with the CD-Bonn
potential based realistic CWG \cite{bro:1}  interaction  used  in
this region. The theoretical results have also been compared with
the  experimental data, wherever available. We have then compared
the two sets of two -body interaction  matrix  elements  to  gain
insight in to the N-N interaction for this n-rich environment.

\section{Shell model results and Discussions}

The  calculations were carried out in the z50n82 model space with
the SMPN \cite{epja} and CWG Hamiltonians \cite{bro:1} using  the
code OXBASH \cite{oxb:1}.

We  started  \cite{138sn:1}  our shell model calculations for the
$^{138}Sn$ nucleus in a slightly truncated  space  excluding  the
intruder  $1i_{13/2}$  neutron  orbital.  This  was  done  on the
observation   of   very   small    (2-5\%)    contributions    of
particle-partitions  involving this orbital in the $^{136-137}Sn$
wavefunctions. No signature of  any  collective  feature  of  the
states  was  found  with  the  increasing  neutron  number in the
N=84-88 isotopes of Sn. Shell model results with the  SMPN  (1+2)
body Hamiltonian were encouraging for these n-rich Sn isotopes. It
was  noted  that  an  estimate \cite{139i:1} of the energy of the
first $2^+$ excited state in  $^{136}Sn$  was  at  600  keV.  The
calculated  result  for  this  was  at  578  keV.  Similarly, the
systematically  estimated  \cite{139i:1}  energy  value  of   the
11/2$^+$  level  in  $^{137}Sn$  was  $\approx$  600  keV and the
calculated value was 550 keV. The estimated \cite{139i:1}  energy
of  the second 5/2$^+$ level in $^{137}Sb$ was also closer to the
calculated result at 773 keV.

With  the  availability  of  the  window  version  of  the OXBASH
\cite{oxb:1}  and  Nushell@MSU  \cite{nush:1}  codes,  it  became
possible  for  us to perform untruncated shell model calculations
including the $\nu 1i_{13/2}$ orbital, for all the A=138  isobars
with  six valence particles in the large basis space of 76 single
particle states mentioned above. The effect of the  inclusion  of
this  intruder  orbital  in the model space has been presented in
Table I. It shows the importance of the full  basis  calculation.
Substantial  changes  in  the  ground  state  binding  energy and
excitation energies of  the  yrast  levels  in  the  spectrum  of
$^{138}Sn$  can be noted. Although the average percentage changes
in the absolute binding energies of the states are $\simeq  2\%$,
the  resultant  effect  on  the  excitation  energy  spectrum  is
dramatic. Wave function compositions of the states for  both  the
calculations  are also given in the Table. Comparison of the wave
functions of the same state of the two  calculations  shows  that
they do not differ much. This is reflected in the B($E2, 0^+_{gs}
\rightarrow  2^+_1$) values of the two calculations, they differ,
but not drastically. The results in  Table  I  are  from  the  SM
calculations with the SMPN interaction.

\subsection{Comparison with systematics}
In  Figs.  1-4  experimental  energy  values  of different states
(yrast 2$^+$, 4$^+$, 6$^+$) for the even-even isotonic series  of
N=70-80,  84-88  and  isotopic series of Z= 50-60 are plotted. As
for $^{136}Sn$ and $^{138}Sn$, experimental level  energy  values
are  not available, we have plotted the predicted values from our
calculations using  SMPN  (Fig.1)  and  CWG  (Fig.2).  Calculated
energies   for   the   first   $2^+$   levels   (E($2^+_1$))   in
$^{136,138}Sn$, using SMPN fit  nicely  in  the  systematics  for
N$>$82  (Fig.1). The E($2^+_1$) values calculated for Sn isotopes
above N=82 decreases with  increasing  neutron  number.  But  the
predicted  E($2^+_1$)  values from the calculations using the CWG
Hamiltonian, deviate sharply  from  the  systematics  for  N$>82$
(Fig.2). However these values agree with the feature observed for
even - even (e-e) Sn isotopes below the $^{132}Sn$ core. For CWG,
the  E($2^+_1$)  excitation  energies  for  all  the even-even Sn
isotopes with N$\geq$ 84 are almost  constant  similar  to  those
below  N=82.  Only  difference is that the bunching of E($2^+_1$)
values now occur at around 750 keV for N$\geq$84 instead of  1200
keV for N$\leq$82.

It  thus  seems that the CWG tbmes still carry the characteristic
features of the not-so-neutron-rich domain below  N=82.  But  the
SMPN  incorporates features of the exotic n-rich domain above the
$^{132}Sn$ core in some of its important  tbmes  appropriate  for
describing  low-lying  states  for nuclei in this region, and its
prediction for low energy eigenvalues seems to be more  reliable,
as  is  evidenced  by  the  trend  in  the  systematics.  Similar
observation is made from the Figs. 3a,b showing the variation  of
the ($E(2^+_1)$ and E($4^+_1$)) of the isotopic series of Z=50 to
60  with  neutron  numbers ranging from 70 to 88. The Figs. 3a,3b
show the results for Sn with  N=86  and  88  from  SMPN  and  CWG
calculations,  respectively.  The  experimental  energies  of  Sn
isotopes stand out from the rest of the nuclei  for  N$<$82  with
nearly  constant values of E$(2^+_1)$ and E($4^+_1$) from N=70 to
80 with a very slow  increasing  trend  with  decreasing  neutron
number.  In  contrast,  for  other isotopic series E$(2_1^+)$ and
E$(4_1^+)$  energies  increase  slowly  with  increasing  neutron
number  and  become  largest  at N=82 and then decrease again for
N$>$82. But for N$>82$, the variation as predicted with both SMPN
and CWG changes dramatically. The SMPN results totally match with
the local systematics. The CWG still retains the  near  constancy
but with much reduced values, with a slight increasing trend with
increasing  neutron  number.  It  seems  that  for N$<$82, the Sn
isotopes were different from the rest due to  their  strong  Z=50
proton shell closed structure. But above $N=82$, the Z=50 closure
becomes  weak and the special feature disappears. This feature is
corroborated by both theoretical results. However CWG results for
Sn show  that  the  variations  of  experimental  E$(2_1^+)$  and
E$(4_1^+)$  with  neutron  number have opposite slopes for N$>$82
compared  to  SMPN  behaviour.  Fig  4  shows  the  variation  of
E$(2^+_1)$,  E($4^+_1$)  and  E($6^+_1$)  for N=84, 86 and 88 for
Z=50-60. The results from SMPN and CWG (for N=86,88) are shown in
the Figure. How the results match   with the systematics  of  the
isotones  and  also  with the trend of N=84 are evident. The SMPN
results fit smoothly in the trend and also  follow  more  closely
the  trend  set  by  $^{134}Sn$. For the CWG results, Z=50 proton
shell closure seems to prevail in a comparatively stronger way.

\subsection{Sn isotopes}
We  now address the issue of the evolution of collectivity in the
even-even $^{134-138}Sn$ isotopes using SMPN \cite{epja} and  CWG
interactions  from  B.A. Brown {\it et al.}\cite{bro:1}. In Table
II,  energies,  wavefunctions  and  B(E2,  $0^+_{gs}  \rightarrow
2^+_1$)  have  been compared for the yrast 0$^+$ and 2$^+$ states
of the three Sn isotopes. As pointed out,  the  E(2$^+_1$)  value
decreases  with  increasing  neutron number for SMPN, but remains
almost constant for  the  CWG  Hamiltonian.  Structure  of  these
states  including  the B(E2,0$^+_{gs} \rightarrow 2^+_1$) values,
as   shown   in   Table    II,    as    well    as    the    R$_4
(=E_{4^+_1}/E/_{2^+_1})$  values suggest that in the calculations
using CWG, the $^{136,138}Sn$ isotopes show a  trend  of  gradual
onset  of collectivity. This is demonstrated by the mild increase
in configuration mixing in the wavefunctions compared to the SMPN
case and an increasing trend in the B(E2,  $0^+_{gs}  \rightarrow
2^+_1$)  values.  The R$_4$ value is also close to 2 (R$_4$=1.78)
for $^{138}Sn$ with CWG. But with SMPN,  results  are  different,
($\nu  2f_{7/2}^n)^{n=2,4,6}$ multiplet configuration is dominant
for the two states in all these isotopes. The value of  R$_4$  is
1.70 for $^{138}Sn$ with SMPN. Although the E$(2^+_1$) values are
relatively  smaller than those predicted by CWG, the structure of
the wavefunctions  and  the  B(E2,  $0^+_{gs}\rightarrow  2^+_1$)
values  show  the  characteristics  of single particle behaviour.
This is a feature already observed experimentally in  the  n-rich
$^{136}Te$   isotope  \cite{rad:1}.  The  anomalously  low  B(E2,
$0^+_{gs}  \rightarrow  2^+_1$)  value  found  in  this   nucleus
compared  to  that  expected  from  the phenomenological modified
Grodzin formula (B(E2,  $0^+_{gs}  \rightarrow  2^+_1$)  $\propto
E_{2^+_1}^{-1}$)   has  drawn  the  attention  of  the  community
\cite{tera:1,dae2002,epja,dae2006}.   Terasaki   {\it   et   al.}
\cite{tera:1}  have  traced  this  anomalous  behaviour in the Te
isotopes to a reduced neutron pairing above the N=82  shell  gap.
This  is  consistent  with  the reduced diagonal n-n tbmes of the
SMPN interaction \cite{dae2002,epja} discussed in Section III  of
this  paper.  It  seems  to  be  a  unique  feature  observed  in
neutron-rich nuclei. In this domain compratively lower  value  of
E($2^+_1$)  does  not  necessarily  lead  to a larger B(E2) value
indicating  collectivity.  $^{136}Te$  probably  was  the   first
nucleus  for  which  this  feature  was  identified \cite{rad:1}.
Interestingly, similar behaviour is also seen near the N=20 shell
closure for neutron-rich Mg and Ne isotopes and also in the odd-A
$^{69-71}Cu$ isotopes \cite{stef:1}.

Despite    being    a    singly    closed   shell   nucleus,   in
$^{32}_{12}Mg_{20}$, the E$(2^+_1)$ is 885 keV, much  lower  than
that  in  the  well deformed $^{24}Mg$ (E$(2^+_1)$=1369 keV). But
the B(E2, $0^+_{gs} \rightarrow  2^+_1$)  in  $^{32}Mg$  is  only
$\simeq$  13  W.u  compared  to  $\simeq$  21  W.u. for $^{24}Mg$
\cite{tra:1}.

It  has  been  pointed out in one of the recent work \cite{khe:1}
that there are indications that the appearance of collectivity in
the yrast levels occurs for nuclei with more than one neutron and
proton in the valence space above the $^{132}Sn$ core. But it  is
still  not very clear how the onset and evolution of collectivity
in these n-rich nuclei exactly depend on the  number  of  valence
neutrons  and protons. So a microscopic investigation using shell
model is indeed necessary and useful.

\subsection{Even-even A=138 isobars}
We   have  studied  the  locus  of  evolution  \cite{138te:1}  of
collectivity as a function of neutron  and  proton  numbers.  Our
observation  regarding this issue of evolution of collectivity in
the low-lying yrast states in all the nuclei in  range  (Z=50-56,
N=82-88)  above  the  $^{132}Sn$  core  has been reported in ref.
\cite{sani:1}. It is found that for nuclei  with  more  than  one
neutron  and  one  proton  in the valence space, mild deformation
appears at N=84, 85, i.e. even before  N=87,  and  deformation  /
collectivity grows with increasing N and Z.

Shell  model  results  for the $^{138}I, ~Cs$ nuclei have already
been discussed  elsewhere  \cite{138i:1,138cs:1}.  These  odd-odd
isobars  have  excitation  spectra,  which  are characteristic of
spherical nuclei.

Here  the  results  for  only the even-even A=138 isobaric series
above  the  $^{132}Sn$  core  have  been  presented.  The  nuclei
considered   are   $^{138}Sn$   ($T_z$=+3.0)   with  6  neutrons,
$^{138}Te$ ($T_z$=+1.0) with 4 neutrons and 2 protons, $^{138}Xe$
($T_z$=-1.0)  with  2  neutrons  and  4  protons   and   finally,
$^{138}Ba$ ($T_z$=-3.0) with 6 valence protons. Untruncated shell
model  calculations  for  6  valence nucleons in this large basis
space usually involve matrices of  large  dimensions,  especially
for  nuclei  having  both  neutrons  and  protons with comparable
numbers or for those having relatively more neutrons (because  of
the  $\nu  1i_{13/2}$  orbital).  The  dimensions of the matrices
involved for different yrast (J$^\pi$,T) - states of a particular
A=138 isobar are shown in Table III. Calculations for  $^{138}Te$
with  2  protons  and  4  neutrons  involve  matrices  of largest
dimensions for a particular spin state.

\subsubsection{Excitation spectra}
For all these A=138 nuclei, except $^{138}Sn$, excitation spectra
are  known.  Calculated  energy  eigenvalues of the level spectra
using both SMPN  and  CWG  Hamiltonians  are  compared  with  the
experimental  data in Figs.(5-8). Calculated energy values of the
levels with the SMPN Hamiltonian are in excellent agreement  with
experimental  ones up to the highest observed level in $^{138}Te$
(Fig.6) and $^{138}Xe$ (Fig.7) (including the non-yrast levels in
$^{138}Xe$, not shown in  the  figure).  The  agreement  for  CWG
results  is  also  very good in both the cases but all the energy
eigenvalues   of   the   excited   states   are    systematically
underpredicted  by about 100-150 keV. The calculations remove the
ambiguity of the spin-parity assignment of the second 8$^+$ state
in $^{138}Te$. From  the  observed  as  well  as  the  calculated
spectra  of  $^{138}Te$  and  the  consideration  of  the $R_4$ =
$(E(4^+_1)/E(2^+_1))$  value,  the   spectrum   appears   to   be
vibrational.  But  closer  inspection  reveals  that  the spacing
between yrast levels (E$_{I+2}$ - E$_I$) increases, although  not
exactly  proportional  to  I (or from the plot of $E_I$ vs I(I+1)
(not shown)). Thus $^{138}Te$ is not a strongly deformed  nucleus
but  lies in the transitional region between spherical $^{136}Te$
and prolate deformed $^{139,140}Te$ \cite{139te:1,138te:1}. It is
interesting to mention here that the $^{137}Te$ spectrum shows  a
kind  of regularity \cite{137te:1}, and the spectrum of $^{139}I$
was  interpreted  as  characteristic  of  a  spherical   vibrator
resulting from a weak interaction of the $^{138}Te$ core with the
odd proton. We may thus term the spectrum of $^{138}Te$ as almost
vibrational.  Spectrum of $^{138}Xe$ is similar to $^{138}Te$ but
the regularity mentioned for $^{138}Te$ is not clearly present in
$^{138}Xe$. However from the consideration  of  the  spectrum  in
$^{139}Xe$  \cite{urb:1} and our recent work \cite{khe:1}, we may
also term the spectrum of $^{138}Xe$ as almost  vibrational.  For
$^{138}Ba$  (Fig.8),  predictions  for  the  energy values of the
level spectra are  also  remarkably  good,  though  the  spacings
between  the  $2^+_2$,  $6^+_2$  and  $2^+_3$ levels are not very
accurately reproduced. But it is remarkable that for  $^{138}Sn$,
where  experimental  level  scheme  is not known, the two results
differ dramatically  (Fig.5).  The  first  three  excited  states
2$^+$,  4$^+$,  6$^+$, as discussed in section A, appear at about
double excitation  energies  in  the  calculation  with  the  CWG
Hamiltonian.  There  is a large energy gap $\simeq$ 2 MeV between
the 6$^+_1$ and 8$^+_1$ states for SMPN calculation. But this gap
is only about 0.8 MeV for CWG results. This energy gap  of  about
0.6-0.8 MeV between $6^+_1$ and $8^+_1$ states also exists in the
experimental  as well as calculated spectra of $^{138}Te,~Xe$ for
both the interactions. In the calculated spectrum of  $^{138}Ba$,
for  both  the  interactions, this gap is $\approx 1.1 MeV$. Thus
the large gap between these states appears  only  with  the  SMPN
Hamiltonian  for  $^{138}Sn$.  These gaps primarily indicate that
the states below and  above  these  gaps  necessarily  belong  to
different  mulitiplets.  Each  state  belong  predominantly  to a
particular muliplet. This can be understood well by  viewing  the
occupancy  plots  in Figs. 9-12. For example, in $^{138}Sn$, $\nu
2f_{7/2}^6$ configuration can couple  to  $J_{max}=6$.  Thus  for
generating the 8$^+$ state a pair of neutrons must be promoted to
a  higher  lying single particle orbital creating the energy gap.
This  gap  for   the   collective   almost   vibrational   nuclei
$^{138}Te,Xe$  also  indicates the dominance of multiplets in the
compositions of the wave functions of these  vibrational  states.
At higher spins the agreement between two calculations improves.

\subsubsection{Binding energies}
The ground state binding energies and spin parities are important
observables  particularly  for application in the nucleosynthesis
calculation. Comparison of binding energies calculated for  these
nuclei using two interactions are shown in the Table IV. The SMPN
results    show    good    agreement   with   experimental   data
\cite{mass2003} for Te, Xe  and  Ba  isotopes.  The  CWG  results
consistently  show  over  binding  for these nuclei, particularly
with increasing proton numbers. But  for  Sn  isotope,  the  SMPN
shows some over binding. This was also noted in our previous work
\cite{epja},  where we have suggested a systematic correction for
neutron - rich species. After introducing that correction for the
Sn isotopes using the binding energy of $^{136}Sn$ from SMPN, the
results  for  $^{138}Sn$  matches  with  those  from   CWG.   The
experimental  binding  energy  value  for $^{138}Sn$ is still not
available. The binding energies for $^{136}Sn$ and $^{138}Te$ are
obtained from systematics  and  therefore  may  have  substantial
errors.  Moreover,  it  is  important to note that CWG predicts a
$2^-$ ground state (B.E=- 55.503 MeV)  instead  of  the  observed
$3^-$    in   $^{138}_{55}Cs_{83}$.   SMPN   correctly   predicts
\cite{138cs:1} the  ground  state  spin-parity  in  it  with  the
binding  energy  value  of  -50.867 MeV, which compares very well
with the measured value of -50.849(17) MeV  \cite{mass2003}.  The
SMPN set of 2-body matrix elements can also reproduce rather well
the  6$^-$ isomer in $^{138}Cs$ \cite{138cs:1}. Interestingly, no
other interaction that we have used, {\it viz.}, KH5082,  CW5082,
CWG  ..  etc.,  usually used in this valence space, can reproduce
the isomeric nature of the  $6^-$  level  with  a  half  life  of
$\approx  $  3  minutes  and  even  the 3$^-$ ground state is not
reproduced in some of these calculations.

For  CWG,  this  deficiency  for nuclei with more valence protons
than neutrons, can perhaps be attributed to the over  binding  of
some  of  the  most important $\pi-\pi$ tbmes (to be discussed in
Section  III).  Similarly,  in  SMPN,  some  important   diagonal
$\nu-\nu$ tbmes still show slight over binding. Better prediction
by  SMPN  is definitely due to the fact that at least some of its
important tbmes contain empirical information appropriate to this
exotic neutron-rich locality. We shall discuss it  further  in  a
following section.

\begin{table}
\caption{Energy  (MeV) spectra and wave functions of yrast levels
of  $^{138}Sn$  in  full  and   truncated   spaces   using   SMPN
interaction.  The  dominant  particle  - partitions are marked as
A($\nu 2f_{7/2}^6$), B ( $\nu 2f_{7/2}^5 1h_{9/2}$) and  C  ($\nu
2f_{7/2}^5   3p_{3/2}$).   The   effect  of  the  intruder  state
$1i_{13/2}$ is denoted by D :($\nu 2f_{7/2}^4 1i_{13/2}^2$).  The
B(E2)   values  are  calculated  for  $0.64e$  value  of  neutron
effective charge.}

\begin{tabular}{ccccc}
\hline
&\multicolumn{2}{c}{Full
space}&\multicolumn{2}{c}{Truncated
space}\\
\cline{2-3} \cline{4-5}
I$^\pi$&Energy& Wavefun. &Energy& Wavefun.  \cr
\hline
0$^+$&0.0         &61\% A & 0.0& 76\% A \cr
     &   (-20.184)&5\% D & (-19.571)&  \cr
\\
2$^+$&0.433&73\%A, 3\%D&0.135& 81\%A\cr
4$^+$&0.738&79\%A, 3\%D&0.407& 86\%A \cr
6$^+$&0.890&80\%A, 3\%D&0.551&86\%A\cr
8$^+$&2.825&64\%B, 4\%D&2.449&77\%C\cr
10$^+$&3.195&77\%B, 2\%D&2.767&81\%B\cr
\hline
\multicolumn{2}{c}{B(E2,$0^+_{gs}\rightarrow 2^+_1$)}
&344.7&&313.9\\
$e^2fm^4$\\
\hline
\end{tabular}
\end{table}

\begin{table}
\setlength{\tabcolsep}{0.001in}
\caption{Comparison  of energies (MeV) and wavefunctions of 0$^+$
and  2$^+$  states  of  $^{134,136,138}Sn$  using  SMPN  and  CWG
interactions.  The  contributions  from  the  dominant $2f_{7/2}$
multiplet ($\nu 2f_{7/2}^n$) configurations are shown, where n is
2,4 and 6 for $^{134,136,138}Sn$, respectively. The B(E2)  values
are calculated for $0.64e$ value of neutron effective charge. The
experimental   2$^+$   energy  for  $^{134}Sn$  is  726  keV  and
B(E2,$0^+_{gs}\rightarrow 2^+_1$) is $\simeq$    295    $e^2fm^4$
\cite{var:1}.}

\begin{tabular}{ccccc}
\hline
&\multicolumn{2}{c}{SMPN \cite{sar:1}}&\multicolumn{2}{c}{Other
\cite{bro:1}}\\
\cline{2-3} \cline{4-5}
I$^\pi$&Energy& Wavefun. &Energy& Wavefun.  \cr
\hline
\multicolumn{2}{l} {$^{134}Sn$}\\
0$^+$&0.0 (fitted)&65\%  & 0.0& 78\%  \cr
     &   (-6.363)&      & (-6.221)&  \cr
2$^+$&~~0.733 (fitted)       &79\%  &0.774& 78\%  \cr
\multicolumn{2}{c}{R$_4 =E_{4^+_1}/E_{2^+_1}$}
&1.46&&1.44\\
\multicolumn{2}{c}{B(E2,$0^+_{gs}\rightarrow 2^+_1$)}
&289.2&&307.8\\
\multicolumn{2}{c}{($e^2fm^4$)}&(1.42 W.u.)&&(1.51 W.u.)\\
\hline
\multicolumn{2}{l} {$^{136}Sn$}\\
0$^+$&0.0         &56\%  & 0.0& 54\%  \cr
     &   (-13.041)&      & (-12.497)&  \cr
2$^+$&0.578       &68\%  &0.733& 49\% \cr
\multicolumn{2}{c}{R$_4 =E_{4^+_1}/E_{2^+_1}$}
&1.53&&1.58\\
\multicolumn{2}{c}{B(E2,$0^+_{gs}\rightarrow 2^+_1$)}
&449.0&&623.5\\
\multicolumn{2}{c}{($e^2fm^4$)}&(2.16 W.u.)&&(3.0 W.u.)\\
\hline
\multicolumn{2}{l}{$^{138}Sn$}\\
0$^+$&0.0         &61\%  & 0.0& 32\%  \cr
     &   (-20.184)&      & (-18.737)&  \cr
2$^+$&0.433       &73\%  &0.761& 24\%  \cr
\multicolumn{2}{c}{R$_4 =E_{4^+_1}/E_{2^+_1}$}
&1.70&&1.78\\
\multicolumn{2}{c}{B(E2,$0^+_{gs}\rightarrow 2^+_1$)}
&344.7&&916.8\\
\multicolumn{2}{c}{($e^2fm^4$)}&(1.63 W.u.)&&(4.33 W.u.)\\
\hline
\end{tabular}
\end{table}

\begin{table}
\caption{Dimensions  of  matrices  involved in different isotopes
for different  yrast states.}

\begin{tabular}{ccccc}
\hline
I$^\pi$&$^{138}Sn$&$^{138}Te$&$^{138}Xe$&$^{138}Ba$\cr
\hline
0$^+$&2411&22960&13416&518 \cr
2$^+$&10713&105489&60955&2134\cr
4$^+$&16321&160243&90982&3053\cr
6$^+$&18351&179054&99241&3170\cr
8$^+$&17271&167325&90114&2715\cr
10$^+$&14284&137121&71324&2000\cr
\hline
\end{tabular}
\end{table}
\begin{table}
\caption{The  binding  enery  (MeV)  results  for different A=138
isobars with SMPN and  CWG.  The  experimental  values  are  also
shown. Experimental errors are shown within parentheses.}

\begin{tabular}{ccccc}
\hline
&\multicolumn{2}{c}{Other \cite{bro:1}
}\\
Nucleus&\multicolumn{3}{c}{Binding Energy (MeV)}\\
&Expt.&\multicolumn{2}{c}{Theoretical}\\
&&SMPN&CWG\cr
\hline
$^{138}Sn$&-& -20.184 &-18.737\\
&-&-18.56$^a$&\\
$^{138}Te$&-36.24(21)&-37.303&-37.692\\
&-&-36.60$^a$&\\
$^{138}Xe$&-48.89( 4)&-49.011&-51.906\\
$^{138}Ba$&-55.44( 1)&-55.266&-62.096\\
\hline
\end{tabular}\\
$^a$  The binding energy values after systematic corrections done
according to Ref. \cite{epja}
\\
\end{table}

\begin{table}
\caption{The extent of configuration mixing involved in different
isotopes  for different states. For each state the numbers in the
parenthesis are (i)  $S$,  sum  of  contributions  from  particle
partitions  each  of  which  is contributing greater than 1.00\%,
(ii) $M$, maximum contribution from  a  single  partition,  (iii)
$N$, total number of partitions contributing in $S$.}

\begin{tabular}{ccccc}
\hline
I$^\pi$&$^{138}_{50}Sn_{88}$&$^{138}_{52}Te_{86}$&$^{138}_{54}Xe_{84}$&$^{138}_{56}
Ba_{82}$\cr
&{$S$, $M$, $N$}&{$S$, $M$, $N$}&{$S$, $M$, $N$}&{$S$, $M$, $N$}\cr
\hline
SMPN\\
0$^+$&~90.7,~60.9,~6&~~69.5,~29.0,~15&~72.5,~19.7,~19&~93.3,~37.6,~10 \cr
2$^+$&~92.4,~73.0,~6&~64.7,~27.1,~15&~65.2,~17.0,~15&~90.7,~38.0,~13 \cr
4$^+$&~93.8,~78.7,~6&~69.6,~29.0,~18&~68.1,~17.8,~16&~91.1,~23.2,~14 \cr
6$^+$&~95.8,~80.0,~7&~71.0,~34.1,~15&~76.1,~24.8,~16&~94.8,~37.0,~13 \cr
8$^+$&~93.0,~64.0,~9&~70.4,~27.2,~15&~69.9,~14.8,~19&~95.3,~31.5,~10 \cr
10$^+$&~94.8,~77.3,~7&~73.5,~27.7,~11&~75.9,~32.2,~16&~98.0,~50.0,~9 \cr
CWG\\
0$^+$&~89.1,~31.9,~12&~~46.2,~11.3,~16&~61.3,~19.3,~16&~96.5,~35.1,~11\cr
2$^+$&~85.5,~23.9,~17&~36.7,~6.4,~17&~59.8,~13.3,~19&~89.8,~40.6,~13 \cr
4$^+$&~86.8,~42.9,~12&~33.9,~5.0,~17&~60.8,~12.8,~21&~92.0,~35.3,~12 \cr
6$^+$&~91.3,~54.9,~9&~42.5,~6.2,~23&~60.1,~14.8,~19&~92.4,~41.0,~9 \cr
8$^+$&~89.7,~43.9,~12&~45.3,~6.1,~26&~57.5,~8.9,~18&~94.3,~45.7,~9 \cr
10$^+$&~92.1,~44.0,~12&~62.1,~9.3,~24&~69.9,~15.7,~21&~95.9,~43.2,~7 \cr
\hline
\end{tabular}
\end{table}
\begin{table}
\caption{The   B(E2,$2^+_1  \rightarrow  0^+_{gs}$) in  $e^2fm^4$  for
different A=138 isobars. The values in Weisskopf units are  also shown
within parentheses.}

\begin{tabular}{ccccc}
\hline
Interactions&$^{138}Sn$&$^{138}Te$&$^{138}Xe$&$^{138}Ba$\cr
            &$T_z$=+3&$T_z$=+1&$T_z$=-1&$T_z$=-3 \cr
SMPN       &68.9(1.6)&484.6(11.5)&638.7(15.1)&530.6(12.5)\cr
CWG        &183.4(4.3)&737.8(17.4)&785.1(18.6)&504.4(11.9)\cr
\hline
\end{tabular}\\
\end{table}

\subsubsection{Wave functions and B(E2) values}
The  structure of the wave functions of different spin states for
the four isobars are compared in Table V and Figs  (9-14).  These
results are obtained using both the SMPN and CWG interaction. The
extent of configuration mixing involved in each yrast state up to
10$^+$  for  these  isotopes  are  compared  in  Table V. We have
tabulated  (i)  $S$,  the  sum  of  contributions  from  particle
partitions  each  of  which  is contributing greater than 1.00\%,
(ii) $M$, maximum contribution from  a  single  partition,  (iii)
$N$,  total  number  of  partitions  contributing  to $S$. Larger
extent of configuration mixing usually leads to larger  deviation
of  S  from  100\%.  Similarly  as mixing increases, M, i.e., the
maximum contribution from a single partition gradually decreases.
Increase in $N$ also indicates larger mixing. From the  table  it
can be found that CWG predicts more configuration mixed structure
for  the  states  in  A=138 isobars compared to that predicted by
SMPN. Figs. 9-12 show the occupancy plots  for  yrast  states  in
each of these nuclei with both interactions. As the figures show,
multiplet  structure  with $\nu 2f_{7/2}^6$ is more clear in SMPN
results (Fig.9) for 0$^+$ to 6$^+$ state in $^{138}Sn$. In Fig.9,
with CWG, increase in contributions  from  $\nu  3p_{3/2}$,  $\nu
2f_{5/2}$  and  $\nu  1h_{9/2}$  orbitals  is  observed.  For Te,
although the mixing increases compared to Sn with SMPN, CWG still
show larger extent of mixing.  Similar  result  is  obtained  for
$^{138}Xe$.  The  B(E2)  peaks at Xe in both SMPN and CWG results
(Table VI). Interestingly, for Ba,  the  two  interactions  give,
except for the binding energy, more or less the same results both
with  respect to the energy spectrum (Fig. 8) as well as the wave
functions and B(E2) values (Table  V,VI  and  Figs.  13,14).  The
experimental   B(E2,   2$^+_1   \rightarrow   0^+_{gs}$)  for  Ba
\cite{tra:1}  is  460  (18)  e$^2fm^4$  compares  well  with  531
e$^2fm^4$  from  SMPN  and  504  e$^2fm^4$ for CWG. The effective
charges used are the standard ones, 1.47$e$ for protons and  0.64
$e$  for neutrons. It can be noted that proton effective charges,
1.37$e$ and 1.41$e$, fit the B(E2,2$^+_1  \rightarrow  0^+_{gs}$)
in  $^{138}Ba$ with SMPN and CWG interactions, respectively. This
value is $\simeq 11$  W.u.,  a  limiting  value  for  vibrational
collectivity.  Although, the B(E2) value for $^{138}Ba$ is in the
range of vibrational collective excitations  and  the  extent  of
configuration  mixing  is  much  larger  than that in $^{138}Sn$,
however,  considering  the  overall  feature  of  the   spectrum,
occupancy  plots  (Fig.12), the wave function structure (Table V,
Fig. 13) and the $R_4$ value of 1.4, it is reasonable to conclude
that the excitations are of single particle  nature.  This  means
N=82 shell closure still prevails in a strong way with increasing
proton  number,  but  the Z = 50 shell closure appears to be much
weaker with the same number of valence  neutrons.  It  is  indeed
satisfactory   to   note   that   the  vibrational  structure  in
$^{138}Te,Xe$  is  excellently  reproduced  in  the  shell  model
calculations.   In   order  to  show  explicitly  the  amount  of
configurations contributing less than 1\% in building  the  yrast
states  in  these  A=138  isobars, we compare in Figs. 13,14, the
wave functions of the  0$^+_1$  states  of  $^{138}Sn,~Ba  $  and
$^{138}Te,  ~Xe$.  From  these $\pi$-chart representations, it is
easy to appreciate how  vigorously  fragmentation  of  parentages
takes  place  as the states go from single particle to collective
nature. The 2$^+_1$ states also show similar  behaviour  for  all
these   isobars.  The  Figs.  9-12  show  time  averaged  nucleon
occupancies of the single  particle  states  for  the  collective
vibrational states in these nuclei. It can be noted that nucleons
are  excited  to  almost all of the single particle states (sps).
For $^{138}Te,Xe$,  any  particular  angular  momentum  state  is
generated  by  the  contributions  from all the various sps (Fig.
14). But, for $^{138}Sn$, (Fig.9)  the  occupancy  plots  clearly
show  the  dominance  of  the contribution from a single orbital,
namely, $\nu 2f_{7/2}$ with both CWG and SMPN interactions. B(E2)
values for $^{138}Te,Xe$, given in Table  VI  are  in  the  range
15-20 W.u., consistent with the characterization of their spectra
as vibrational.

So  these  comparisons  clearly  indicate  that  as  far  as  the
proton-proton two-body matrix elements are concerned,  these  two
interactions  have  over all similar performances, except for the
over binding of the $\pi-\pi$ tbmes in CWG, affecting the binding
energies (with respect to the $^{132}Sn$  core)  of  the  states.
Moreover,  it  can  be  noted  from  Table VI, that the predicted
B(E2,2$^+_1  \rightarrow   0^+_{gs}$)   values   in   $^{138}Sn$,
$^{138}Te$  and  $^{138}Xe$  are  somewhat  larger  with  the CWG
interaction.

\section{Comparison of tbmes : empirical and realistic interactions}

Spectroscopic   data  for  very  neutron-rich  nuclei  above  the
$^{132}Sn$ core are being available recently. But available  data
are  not sufficient yet to construct a fully empirical (1+2) body
Hamiltonian. It is interesting to point out that the construction
of  empirical  Hamiltonians  for  the  n-rich   environments   in
different  mass  regions  might  be  important  for  checking the
realistic  interactions  built  from  the  fundamental  bare  N-N
interaction.  Comparison  of  the  predictability  of a realistic
Hamiltonian over a model space, as well as its  correlation  with
an empirical Hamiltonian with good predictive power over the same
model  space  can  help  understanding  how realistically various
nucleon-nucleon correlations have been taken  in  to  account  in
obtaining the realistic Hamiltonian.

For  microscopic  understanding  of  the  available  data  on the
structure of these very n-rich nuclei, one finds spherical  shell
model  calculations  very suitable for this region and thus needs
appropriate nuclear  Hamiltonian.  Available  KH5082  and  CW5082
Hamiltonians and very recently effective interaction derived from
CD-Bonn   potential   are   being   used.  In  our  earlier  work
\cite{sar:1}  we  have  shown  that  CW5082   Hamiltonian   works
reasonably  well  for  N$<$84  isotones  but  fails for N$\geq$84
isotones.

Our  work  in  Ref. \cite{epja}, following the method of Chou and
Warburton \cite{cw:1} is perhaps  the  first  somewhat  elaborate
empirical  approach  to  improve  upon  the  existing Hamiltonian
appropriate for this particular region.  We  used  the  available
experimental  data  for  the A=134 Sn, Sb and Te nuclei available
then. As mentioned above, our earlier studies showed that  CW5082
is  reasonable for N$<$84. So we started with CW5082 and modified
it as far as  practicable  with  the  limited  experimental  data
available at that point of time. The motivation for these changes
was  also  clearly  stated  in  Ref.  \cite{epja}. All the single
particle energies (spes) of the valence orbitals  in  the  CW5082
were  replaced by experimentally determined ones, except the $\pi
3s_{1/2}$ and $\nu 1i_{13/2}$ spes (Table VII).  These  two  spes
were obtained as stated in Ref. \cite{epja}. Out of 2101 two-body
matrix  elements,  26  tbmes  of  CW5082 interaction were changed
using the data on the binding energies of the ground  states  and
low-lying  spectra  of  the A=134 isobars of Sn, Sb and Te nuclei
only and not any other nuclei above the core. It should be  noted
that    these   A=134   nuclei   provide   information   on   the
neutron-neutron  ($\nu-\nu$),  proton-neutron   ($\pi-\nu$)   and
proton-proton  ($\pi-\pi$)  tbmes,  respectively. Modified CW5082
was named as SMPN and have been used for shell model calculations
for almost all neutron-rich nuclei above $^{132}Sn$ core  in  the
range   134$\leq$A$\leq$138.   Some  of  the  results  have  been
presented  in  \cite{epja,136i:1,138i:1,138cs:1}.  It  should  be
mentioned  that  the  new  interaction  predict  better  results,
compared to CW5082 and other  interactions  being  used  in  this
region,  for  N$<$84  nuclei  too. The relevant spes and tbmes of
CW5082, CWG \cite{bro:1} and SMPN are compared in Tables VII  and
VIII, respectively.

It was already pointed out by Chou and Warburton \cite{cw:1} that
a  glaring  deficiency  in  Kuo-Herling interaction, which is the
basic ingredient of CW5082, is that the neutron-neutron J=0 tbmes
are too attractive. So for  an  approximate  adjustment  of  this
defect  those  tbmes  were multiplied by 0.6 in KH5082 and CW5082
\cite{cw:1}. Six diagonal $\nu-\nu$ tbmes  with  (J$^\pi$,  T)  =
(0$^+$,1)  of  SMPN are further reduced by about 50\% compared to
those of CW5082. The reduction  of  these  tbmes,  the  so-called
pairing terms were obtained \cite{dae2002,epja} by adjusting them
to  reproduce  the  binding  energy  of $^{134}Sn$. Corresponding
tbmes of CWG are relatively overbound.

The  neutron-neutron  J=0  tbmes,  the  pairing terms, are indeed
different from those expected from standard pairing  interaction.
But  in  view of the consistently improved predictability of SMPN
for the binding energies, level spectra and  other  spectroscopic
properties  for  almost  all  the  n-rich  nuclei  in  the  range
134$\leq$A$\leq$138  as  delineated  in  the  present  work   and
elsewhere \cite{epja,136i:1,138i:1,138cs:1}, it may be reasonable
to  conclude that the $\nu- \nu$ pairing terms are indeed reduced
in this neutron -rich environment. Moreover, slight over  binding
of  the ground states of Sn isotopes above $^{132}Sn$ as obtained
with the SMPN Hamiltonian suggests  further  reduction  of  these
matrix   elements,   which   may   be  due  to  increasing  A  or
uncertainties  in  some  of  the  spes.  This  reduction  of  the
$\nu-\nu$  pairing  terms  observed  \cite{dae2002,epja} from the
shell  model  applications  is   consistent   with   the   recent
observations \cite{tera:1,ritu:1}.

Three    other    modified    diagonal   $\nu-\nu$   tbmes   with
($J^\pi$,T)=(2$^+$, 4$^+$, 6$^+$,1) are also given in the  table.
Tbmes $\langle 2f_{7/2}^2 |V|2f_{7/2}^2\rangle ^{J=2,4}$ have the
same  phases for all the three interactions, but their magnitudes
in SMPN are larger than the corresponding values  in  CW5082  and
CWG.  This  is  because we have assumed purer multiplet structure
for 2$^+$ to 6$^+$  and  also  for  $8^+$  of  $^{134}Sn$,  while
adjusting  the  tbmes to reproduce the experimental level scheme.
Diagonal $\langle 2f_{7/2}^2 |V|2f_{7/2}^2\rangle ^{J=6}$ of SMPN
is out of phase with that of CW5082 and CWG. The  other  modified
$\nu-\nu$ tbme $\langle 1h_{9/2}^22f_{7/2}
|V|1h_{9/2}^22f_{7/2}\rangle  ^{J=8}$  also  has  somewhat larger
value in SMPN.

Proton  -  proton  diagonal  tbmes  of  the CWG with J=0,2,4 show
overbinding compared to other two interactions. The CWG tbme with
J=6 has much smaller magnitude compared to that for the other two
interactions. Comparison of the binding energies and spectra  for
$^{138}Te$, $^{138}Xe$ and $^{138}Ba$ clearly suggests that these
most  important  $\pi-\pi $ tbmes are indeed overbound in case of
the CWG interaction.

All the neutron-proton ($\pi-\nu$) tbmes in the T=0 channel given
in  the  table  have the same phases in these three interactions.
But the $\pi-\nu$ tbmes of the SMPN interaction are found  to  be
stronger.

\begin{table}
\caption{Single  particle  energies  used in SMPN \cite{epja},
CWG \cite{bro:1} and CW5082 \cite{cw:1} Hamiltonians.}
\begin{tabular}{cccccccc}
\hline
 State     &\multispan{3}    \hfil    Energy    (MeV)\hfil & State
&\multispan{3}
\hfil    Energy    (MeV)\hfil    \\
                  & SMPN   & CWG&CW5082&& SMPN   & CWG&CW5082     \\
\hline
$\pi 1g_{7/2}$ & -9.66290& -9.66300&-9.59581& $\nu 1h_{9/2}$& -0.89400&-0.89440&-
0.89500\cr
$\pi 2d_{5/2}$ & -8.70000& -8.70050&-8.67553& $\nu 2f_{7/2}$& -2.45500&-2.45530&-
2.38000\ \cr
$\pi 2d_{3/2}$ & -7.22300& -7.22328&-6.95519&$\nu 2f_{5/2}$& -0.45000&-0.45070&-
0.89000 \cr
$\pi 3s_{1/2}$ & -7.32300& -6.96570&-6.92783&$\nu 3p_{3/2}$& -1.60100&-1.60160&-
1.62500\cr
$\pi 1h_{11/2}$& -6.87000& -6.87141&-6.83792&$\nu 3p_{1/2}$& -0.79900&-0.79960&-
1.16000\cr
&&&&$\nu 1i_{13/2}$&0.25000&0.239700&-0.29000\\
\hline
\end{tabular}
\end{table}
\begin{table}
\vspace{-2cm}
\caption{The   relevant   tbmes   of   CW5082   \cite{cw:1},  CWG
\cite{bro:1} and SMPN \cite{epja} have been compared. The valence
orbitals  have  been  enumerated  according  to   the   following
convention:     PROTONS     :    1$g_{7/2}$(1),    2$d_{5/2}$(2),
2$d_{3/2}$(3),     3$s_{1/2}$(4),    1$h_{11/2}$(5);    NEUTRONS:
1$h_{9/2}$(6), 2$f_{7/2}$(7), 2$f_{5/2}$(8), 3$p_{3/2}$(9),
3$p_{1/2}$(10), 1$i_{13/2}$(11)}

\begin{tabular}{ccccccccc}
\hline
\multicolumn{4}{c}{Indices}&J&T&\multicolumn{3}{c}{Hamiltonians}\\
&&&&&& CW5082&          CWG &   SMPN\\
\hline
\multicolumn{7}{l}{n-p tbmes}\\
1 & 7 & 1 & 7  &0 &0&   -0.6778&        -0.7922&    -0.7355\\
1&  7&  1&  7&   1&  0&   -0.336&         -0.5390 &   -0.743\\
1& 7 &1 &7 & 2 &0 & -0.29336119 &  -0.3739   &-0.4067\\
1& 7 &1& 7 & 3 &0 & -0.30348513 &  -0.2385   &-0.354\\
1& 7 &1& 7 & 4 &0 & -0.09123172 &  -0.1290   &-0.227\\
1& 7 &1& 7 & 7 &0 & -0.42997605 &  -0.5440   &-0.535\\
1& 6 &1& 6 & 8 &0 & -0.68749624 &  -0.9491   &-1.037\\
1&11 &1&11 &10 &0 & -0.61499959 &  -0.7795   &-0.764\\
5& 7 &5& 7 & 9 &0 & -0.57939130 &  -0.5890   &-1.058\\
5&11 &5&11 &10 &0 & -0.08622793 &  -0.2972   &-1.942\\
5&11 &5&11 &11 &0 & -0.12218534 &  -0.0354   &-1.611\\
5&11 &5&11 &12 &0 & -0.75382543 &  -1.1495   &-1.519\\
\multicolumn{7}{l}{p-p tbmes}\\
1& 1& 1& 1&  0& 1  &-0.66390    &  -0.9972   &-0.56100\\
1& 1& 1& 1&  2& 1  & 0.16450    &  -0.2838   & 0.2050\\
1& 1& 1& 1&  4& 1  & 0.38240    &  -0.0257   & 0.5682\\
1& 1& 1& 1&  6& 1  & 0.43640    &   0.1008   & 0.5120\\
\multicolumn{7}{l}{n-n tbmes:
pairing terms}\\
6 &6 &6 &6  &0 &1  &-0.61197406 &  -1.2944   &-0.293747544\\
7 &7 &7 &7  &0 &1  &-0.48571584 &  -0.6718   &-0.233143598\\
8 &8 &8 &8  &0 &1  &-0.27602252 &  -0.4515   &-0.132490814\\
9 &9 &9 &9  &0 &1  &-0.37947276 &  -0.5884   &-0.182146922\\
10&10&10&10 &0 &1  &-0.10030835 &  -0.1606   &-0.0481480062\\
11&11&11&11 &0 & 1 &-0.70925683 &  -0.9751   &-0.340443283\\
\multicolumn{7}{l}{n-n tbmes}\\
7 &7 &7 &7  &2 &1 & -0.31314358 &  -0.2983   &-0.453\\
7 &7 &7 &7  &4 &1 & -0.05632162 &  -0.0909   &-0.29\\
7 &7 &7 &7  &6 &1 &  0.05457612 &   0.0011   &-0.162\\
6 &7 &6 &7  &8 &1 & -0.25914928 &  -0.3974   &-0.495\\
\hline
\end{tabular}
\end{table}
\section{Summary and Conclusions}
The  neutron-  rich $^{132}$Sn region is an interesting domain to
study the N-N interaction in the  neutron-rich  environment.  Our
comparison   of  the  calculated  results  for  the  $E_{2^+_1}$,
$E_{4^+_1}$  and  $E_{6^+_1}$  level  energies  with  the   local
systematic  trends  reveal  the  evolution of the shell structure
below and above the Z = 50,  N  =  82  shell  closure.  Both  the
interactions,  SMPN  and CWG are quite capable of reproducing the
existing experimental data for A=138 isobars, namely  $^{138}Te$,
$^{138}Xe$   and   $^{138}Ba$.  But  the  predicted  spectra  for
$^{138}Sn$ using these two interactions differ significantly. The
SMPN  results  fit  nicely  to  the  systematics  of  the  nuclei
(Z=52-60)  above  $^{132}Sn$ core. For neutron number $< 82$, the
$Sn$ (Z=50) isotopes have a special status due to its semi- shell
closed structure. But for increasing neutron number ($>82$),  the
Z=50  shell closure weakens and $Sn$ isotopes behave very similar
to the other Z=52-60 nuclei. But with CWG, the weakening  is  not
much   dramatic,   Sn   isotopes   (N=86,88)  still  retain  some
speciality, although to a lesser extent. Which picture is correct
and  represent  the  real  situation  can  only  be  resolved  by
experimental  data.  Comparing  the  structure  of  the states in
$^{138}Ba$ with six valence protons with that of  the  $^{138}Sn$
with the same number of neutrons, it is found that the protons in
this valence space are more efficient in generating configuration
mixing.   This  is  essentially  due  to  close  spacing  between
$\pi(gds)$  single  particle  orbitals  that  permits   $\pi-\pi$
residual  interaction to scatter easily protons to various single
particle  states.  This  comparison  also  indicates   that   the
increasing  neutron  number  weakens  Z=50  shell  closure  in  a
stronger way than an equal number of protons can do to  the  N=82
shell closure.

The  results  for  the spectra and binding energies of the ground
states of $^{138}Te$, $^{138}Xe$ and  $^{138}Ba$,  when  compared
more  precisely  show that SMPN predictions are more close to the
experimental  data.  This  better  predictability  of  the   SMPN
interaction  over the CWG may be attributed to the fact that SMPN
incorporates in some of its important tbmes, the special features
of the N-N interaction in the exotic n-rich environment above the
$^{132}$Sn  core.  This  perhaps  suggests  for   exotic   n-rich
environments in general, the essentiality of incorporation of the
local  spectroscopic  information in obtaining tbmes of the shell
model Hamiltonian for better predictability towards dripline.

Our   results   using  SMPN  interaction  clearly  rule  out  the
possibility of onset  of  collectivity  in  the  low-lying  yrast
states  of even Sn isotopes around N=88, whereas the results with
the CWG interaction possibly shows an indication of gradual onset
of  collectivity.  However,  both  the   interactions   reproduce
remarkably   well   the   collective   vibrational   spectra   in
$^{138}$Te,Xe, with the equal number of np-pairs. For nuclei with
more than one neutron and  proton  in  the  valence  space,  mild
deformation  appears  even before N=87, at N=84,85 ($^{137}$I and
$^{137}$Te) and deformation/collectivity grows with increasing  N
and  Z.  In  SMPN, tbmes scaled from non exotic $^{208}Pb$ region
have been changed by using spectroscopic information  of  A  =134
isobars  of Sn,Sb and Te only. Particularly, the diagonal J=0 n-n
tbmes, the so called pairing terms, had to be reduced by a factor
of 0.288.  In  CWG  too,  n-n  tbmes  have  been  reduced.  These
reduction  of  n-n  pairing  interaction is consistent with other
recent results as discussed above and seems  to  be  generic  for
exotic n- rich environments. Comparison of the tbmes of the three
interactions  enhances  the  belief  that  the N-N interaction is
indeed different in nuclei in the very  n-rich  environment  than
that near the line of stability.

\section{Acknowledgments}
The  authors  thank  Prof.  Waldek  Urban  for  many  stimulating
discussions on the issues in this mass region. Special thanks are
due to Prof. B.A. Brown for his help in providing us  his  OXBASH
(Windows Version) and the Nushell@MSU code. \\

\begin{figure}[h]
\vspace{12.5cm}
\includegraphics{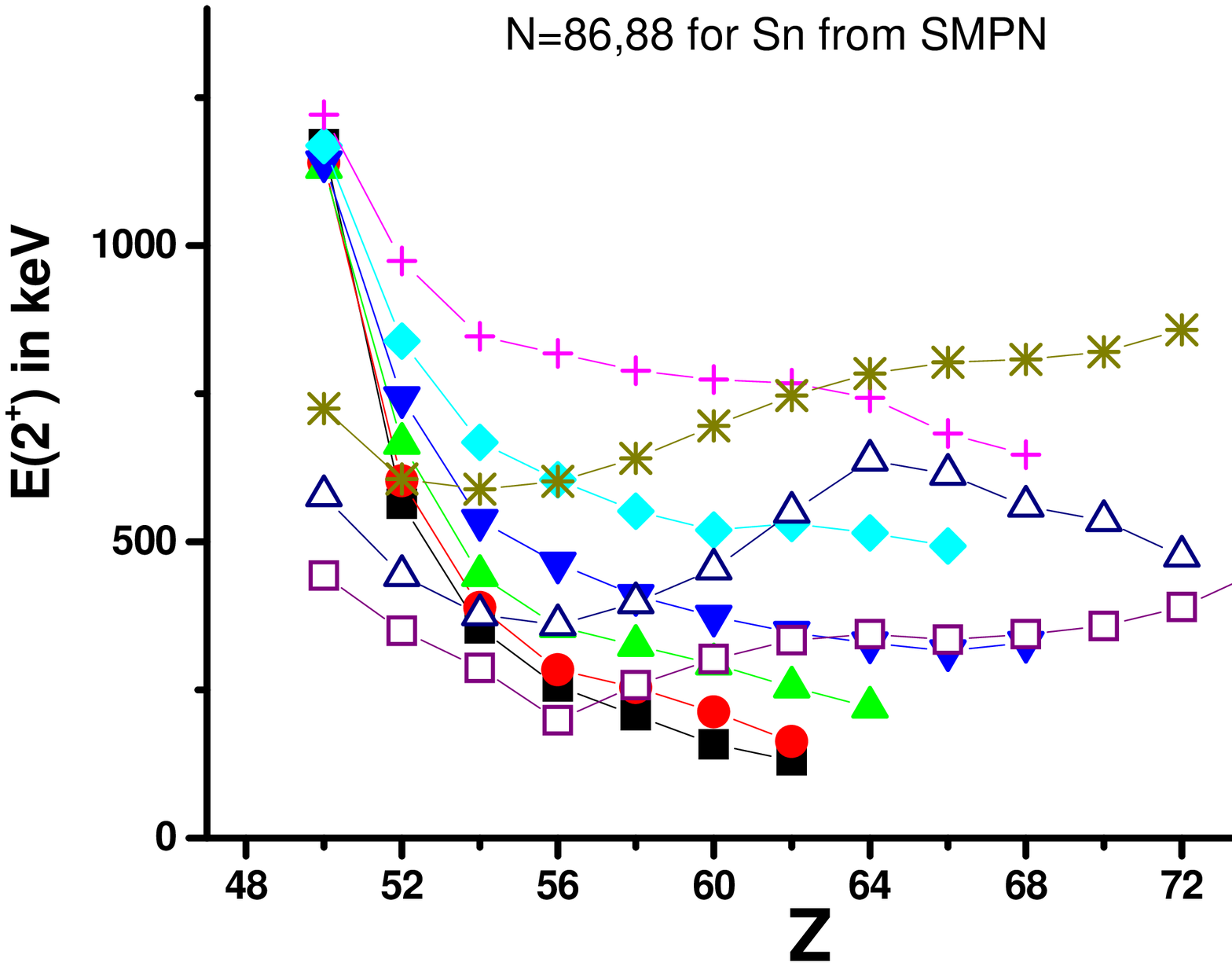}
\vspace{1.5cm}
\caption{\label{fig:fig1}  (Color online) Variation of $E(2^+_1)$
energies  in  the   isotonic   series.   Predicted   values   for
$^{136,138}Sn$  using  SMPN  are  shown  in the figure. The other
values are from experiments \cite{nndc}.}

\end{figure}
\begin{figure}[h]
\vspace{12.5cm}
\includegraphics{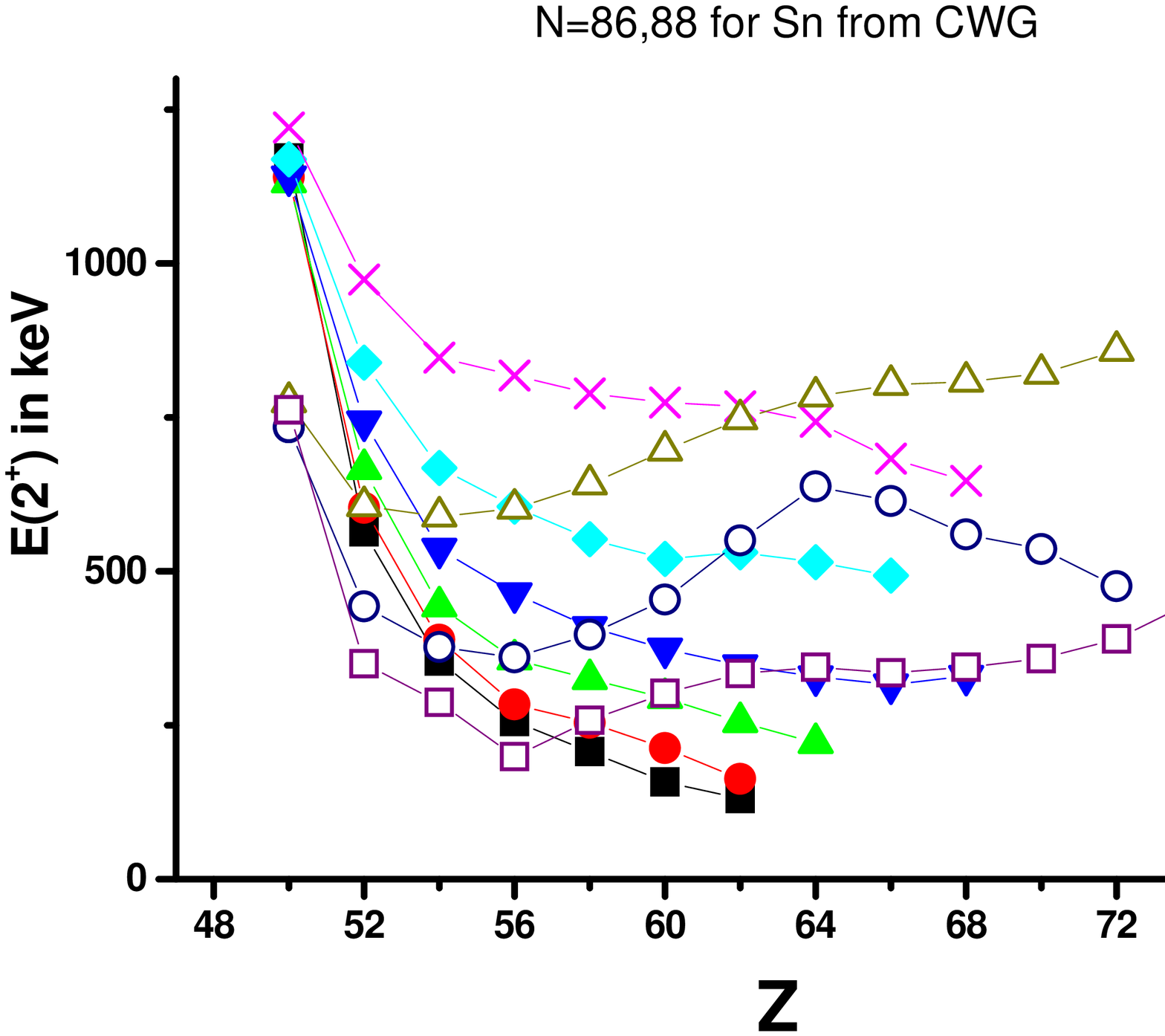}
\vspace{1.5cm}
\caption{\label{fig:fig2}  (Color online) Variation of $E(2^+_1)$
energies  in  the   isotonic   series.   Predicted   values   for
$^{136,138}Sn$  using  CWG  are  shown  in  the figure. The other
values are from experiments \cite{nndc}.}

\end{figure}
\begin{figure}[h]
\vspace{12.5cm}
\includegraphics{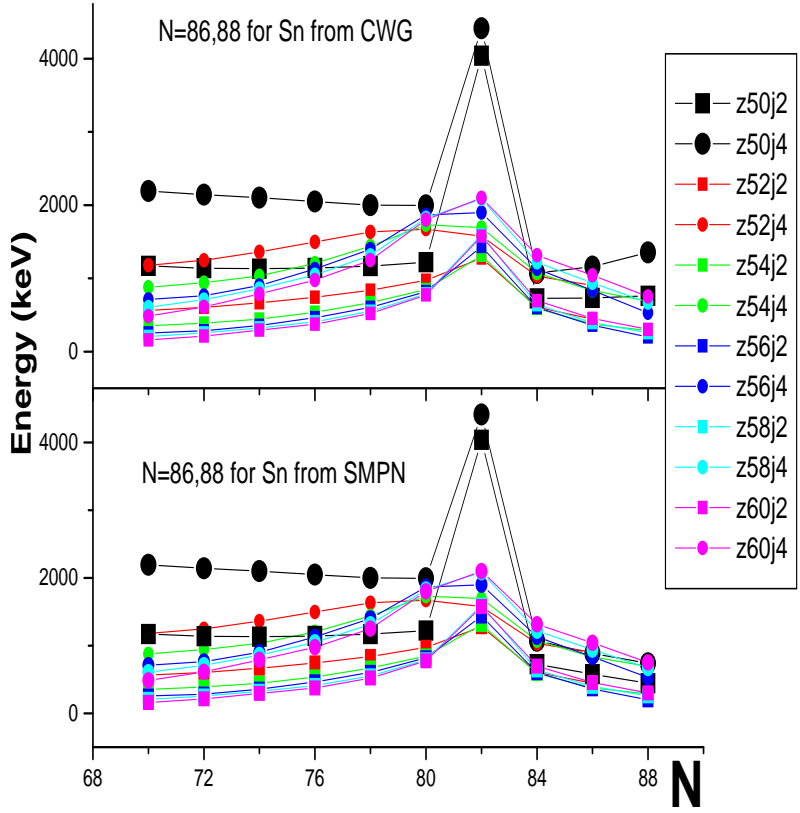}
\vspace{3.5cm}
\caption{\label{fig:fig3}   (Color   online)   Variation  of  the
($E(2^+_1)$ and E($4^+_1$)) of the isotopic series of Z=50 to  60
with  neutron  numbers  ranging from 70 to 88 is shown. Predicted
values for $^{136,138}Sn$ using CWG are shown  in  upper  figure,
while those using SMPN are in the lower one. The other values are
from experiments \cite{nndc}.}

\end{figure}
\begin{figure}[h]
\vspace{12.5cm}
\includegraphics{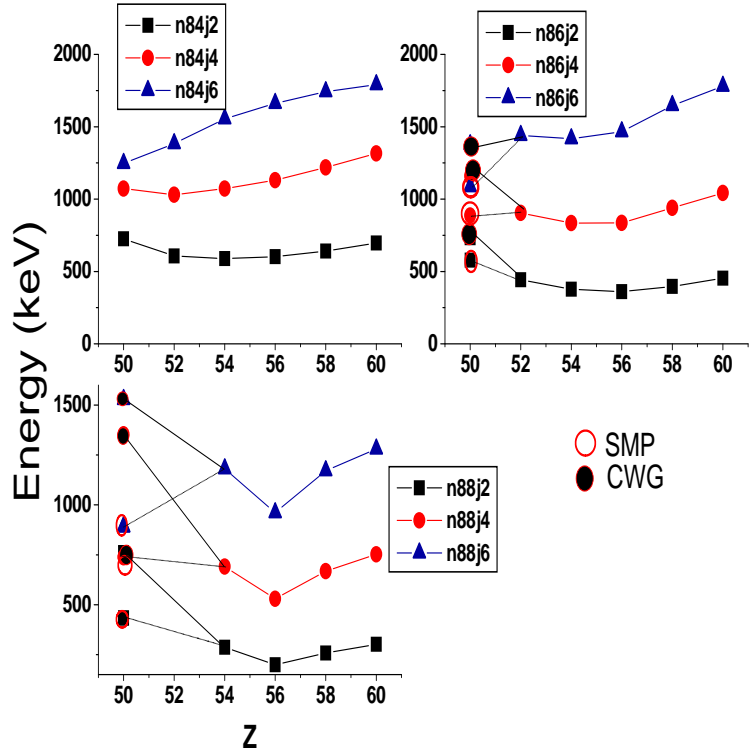}
\vspace{2.5cm}
\caption{\label{fig:fig4}(Color  online) Variation of E$(2^+_1)$,
E($4^+_1$) and E($6^+_1$) for N=84, 86  and  88  for  Z=50-60  is
shown.  The  results from SMPN and CWG (for N=86,88) are shown in
the Figure. The other values are from experiments \cite{nndc}.}

\end{figure}
\noindent
\begin{figure}
\vspace{8.5cm}
\includegraphics{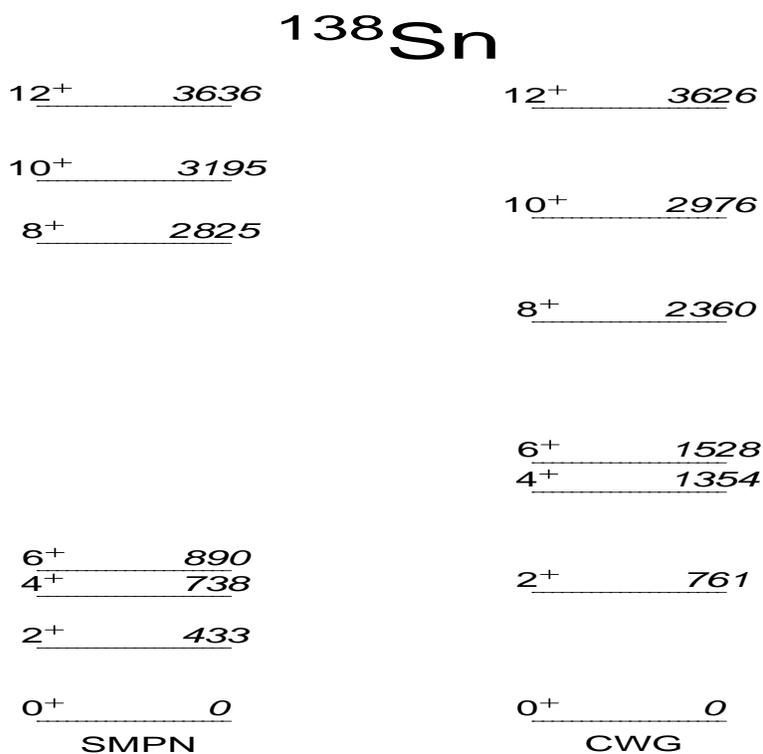}
\vspace{1.5cm}
\caption{\label{fig:fig5} The calculated excitation spectra using
SMPN   and   CWG  interactions.}
\end{figure}

\begin{figure}[t]
\vspace{8.5cm}
\includegraphics{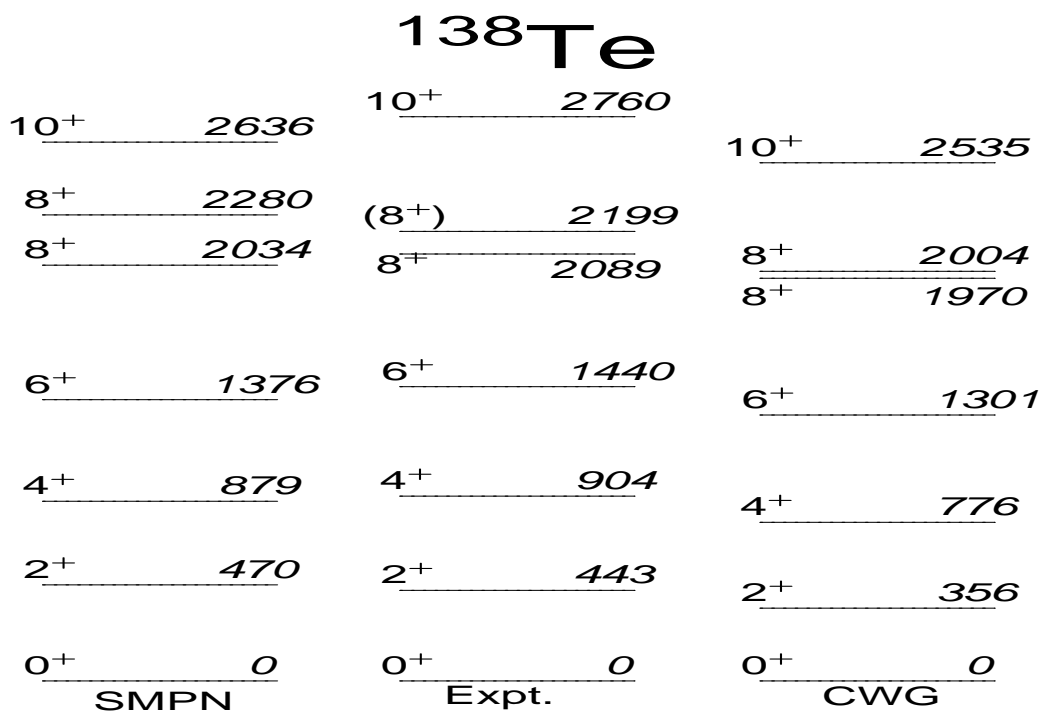}
\vspace{1.5cm}
\caption{\label{fig:fig6}   Comparison   of   experimental    and
calculated  excitation  spectra using SMPN and CWG interactions.}
\end{figure}

\begin{figure}[t]
\vspace{8.5cm}
\includegraphics{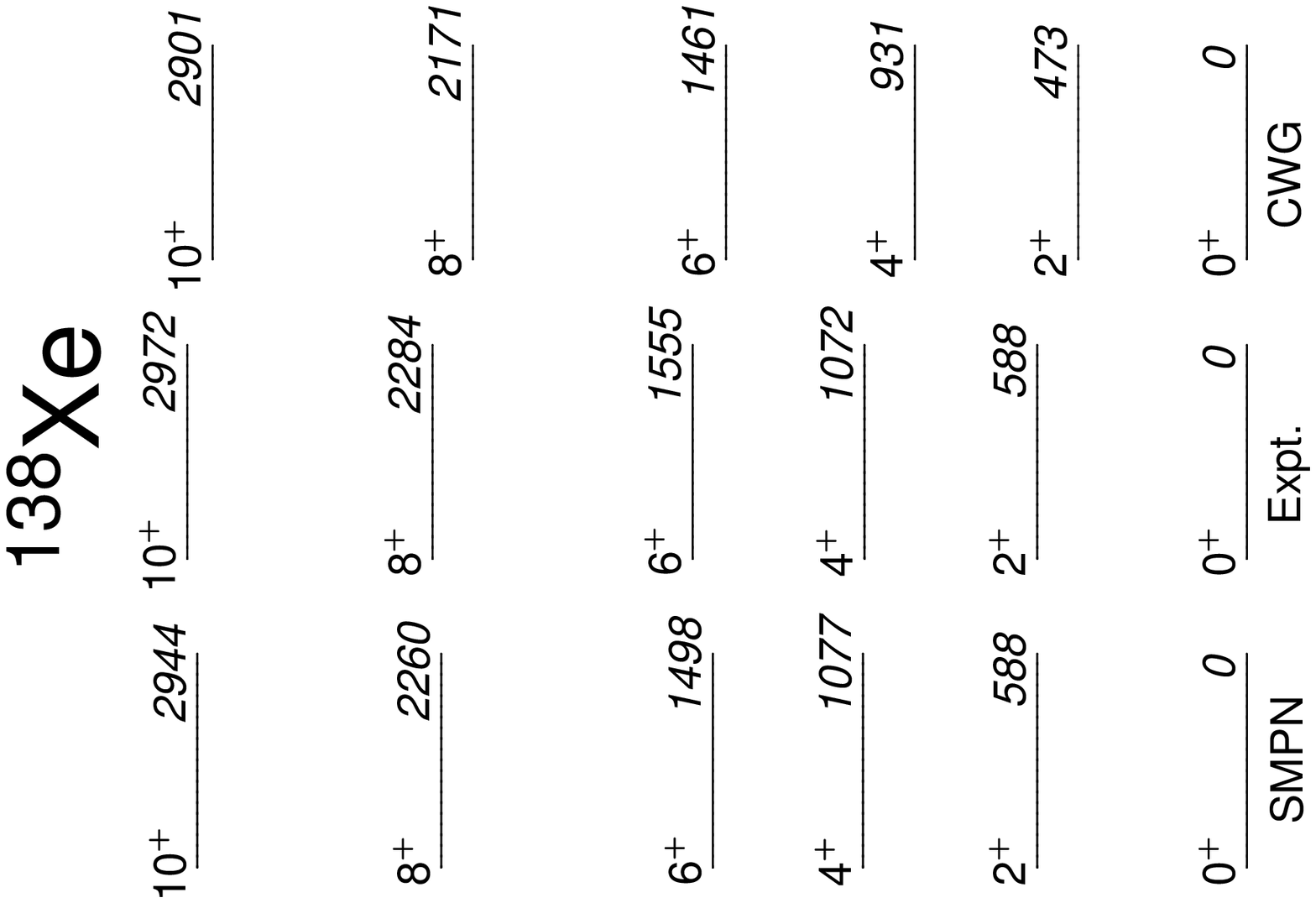}
\vspace{1.5cm}
\caption{\label{fig:fig7}Same      as     Fig.6}
\end{figure}

\begin{figure}[b]
\vspace{8.5cm}
\includegraphics{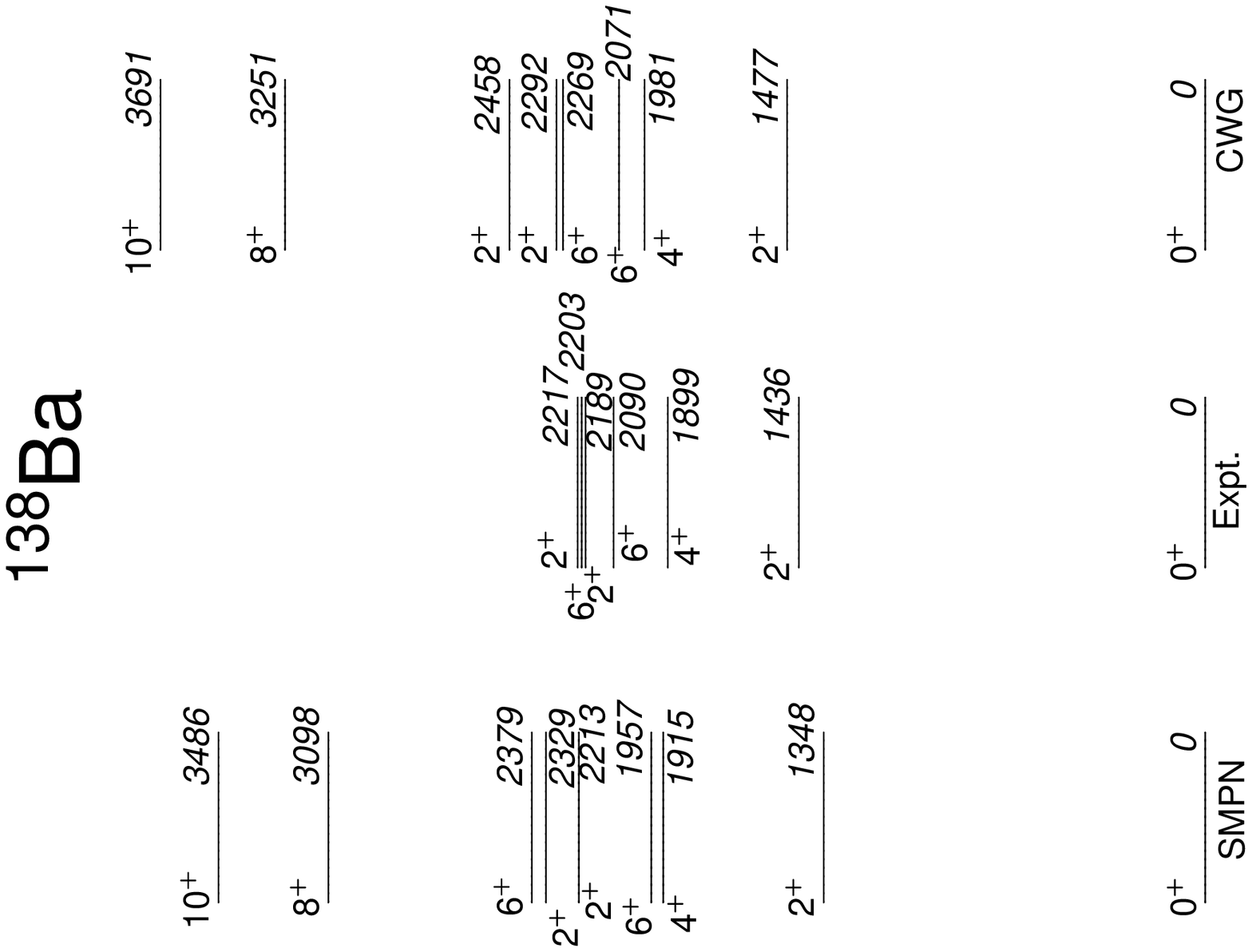}
\vspace{1.5cm}
\caption{\label{fig:fig8}Same     as     Fig.6}
\end{figure}

\begin{figure}[h]
\vspace{12.5cm}
\includegraphics{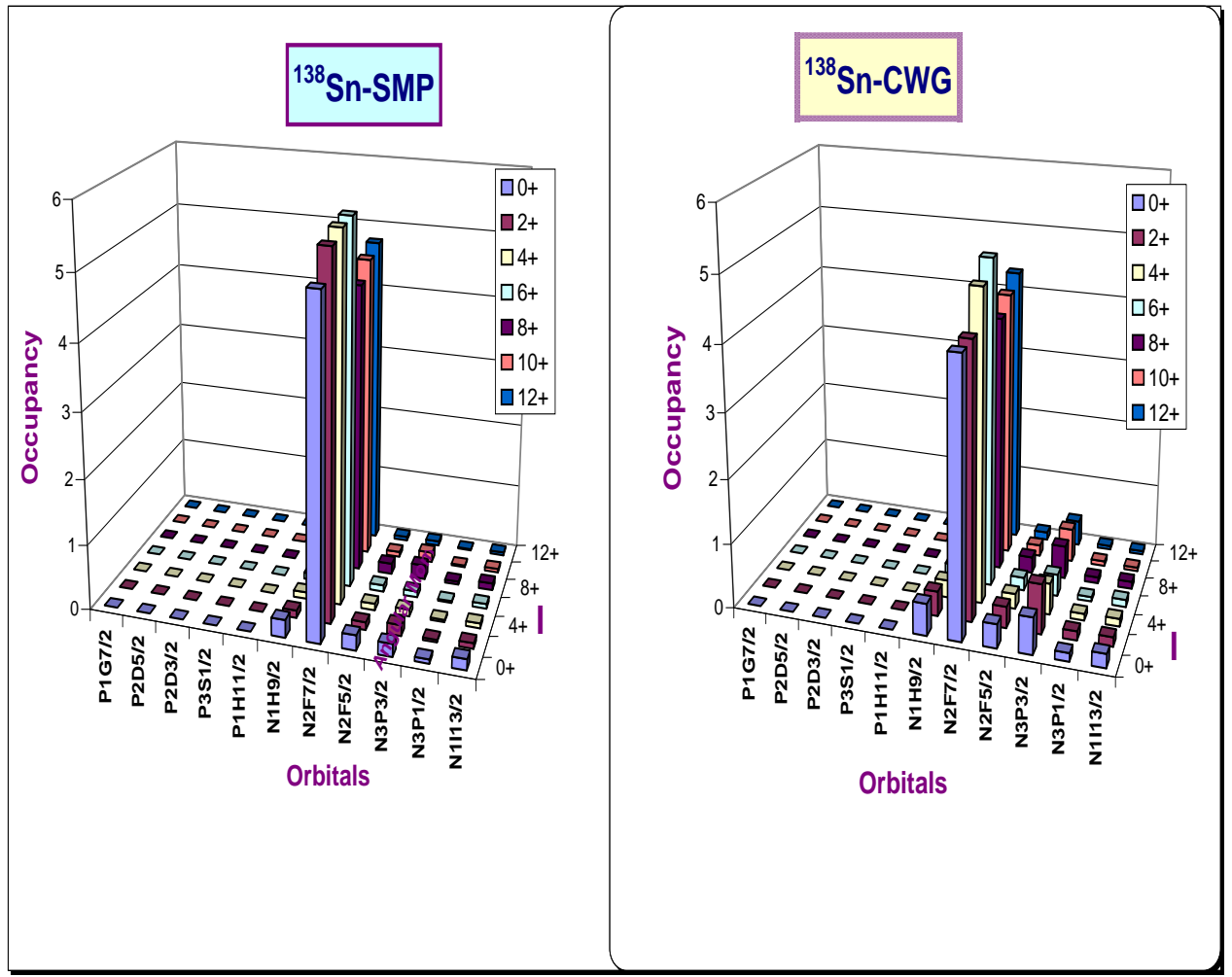}
\vspace{1.5cm}
\caption{\label{fig:fig9}  (Color  online) The number of neutrons
occupying the orbitals indicated along the x-axis are plotted for
different angular momentum states of this  nucleus.  The  results
using two interactions are indicated in the figure.}

\end{figure}
\begin{figure}[h]
\vspace{12.5cm}
\includegraphics{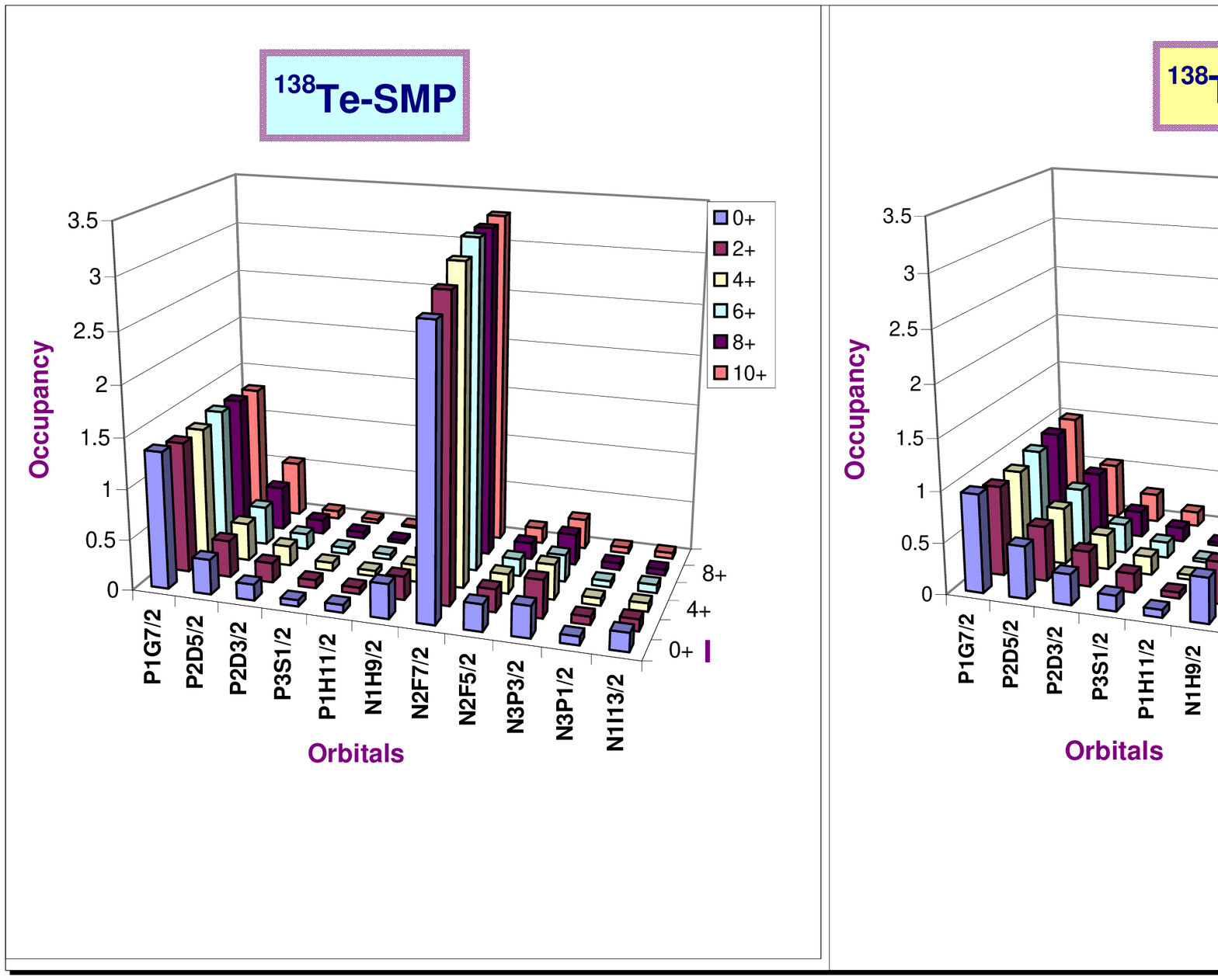}
\vspace{1.5cm}
\caption{\label{fig:fig10}  (Color online) The plot of proton and
neutron occupancies of different orbitals. See also  the  caption
of Fig.9}

\end{figure}

\begin{figure}[h]
\vspace{12.5cm}
\includegraphics{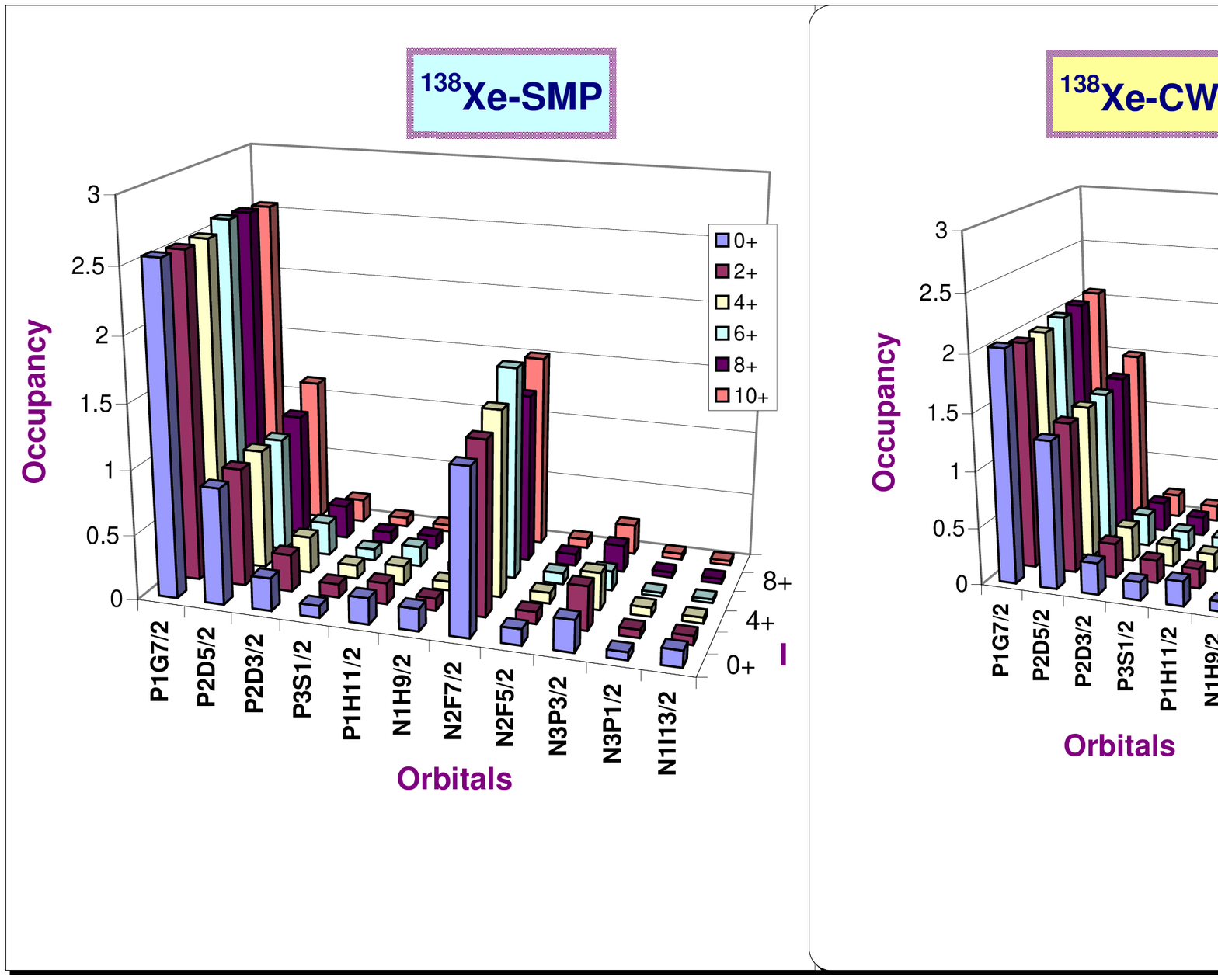}
\vspace{1.5cm}
\caption{\label{fig:fig11}  (Color online) The plot of proton and
neutron occupancies of different orbitals. See also  the  caption
of Fig.9}

\end{figure}
\begin{figure}[h]
\vspace{12.5cm}
\includegraphics{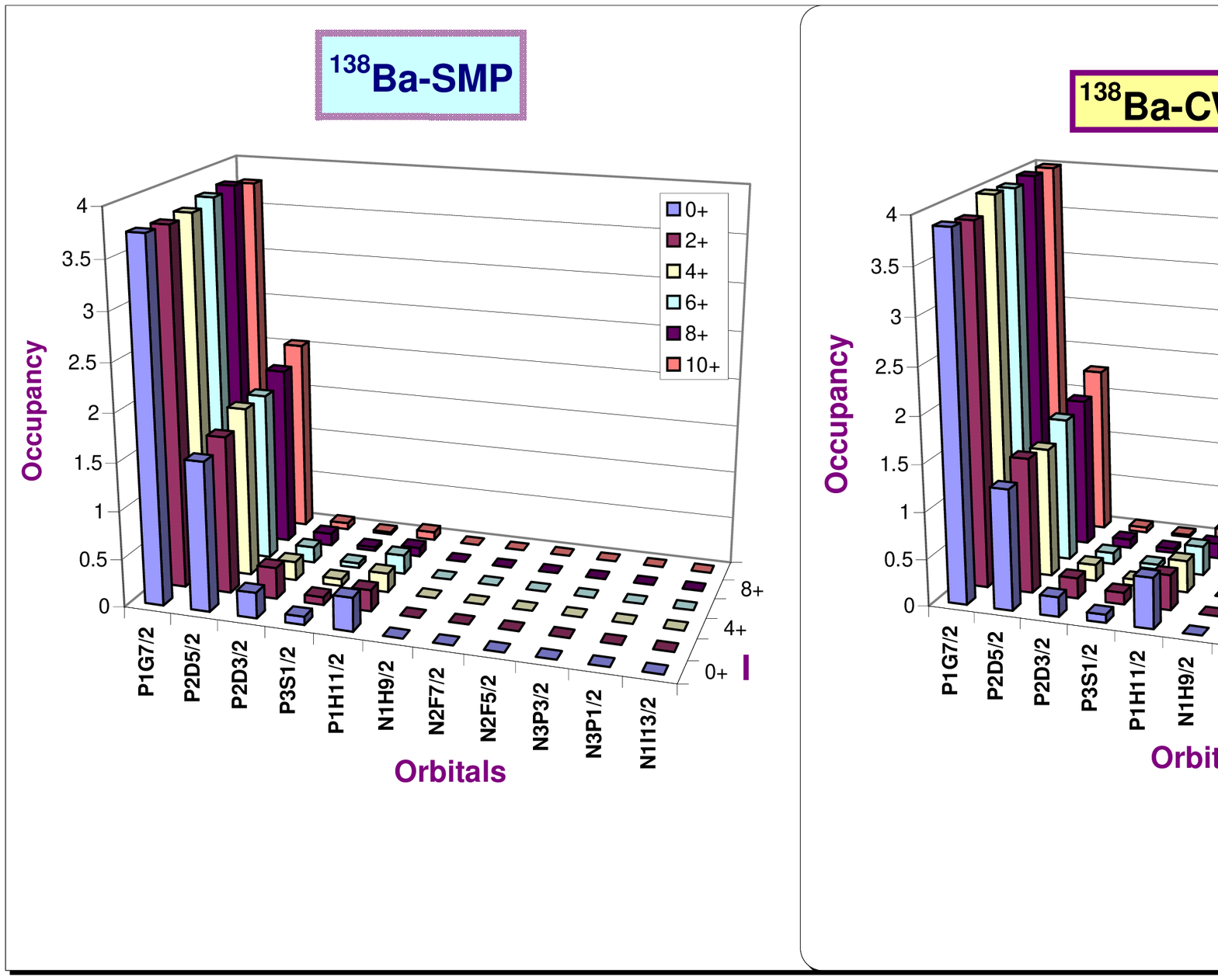}
\vspace{1.5cm}
\caption{\label{fig:fig12}(Color   online)  The  plot  of  proton
occupancies of different orbitals. See also the caption of Fig.9}

\end{figure}
\begin{figure}[h]
\vspace{12.5cm}
\includegraphics{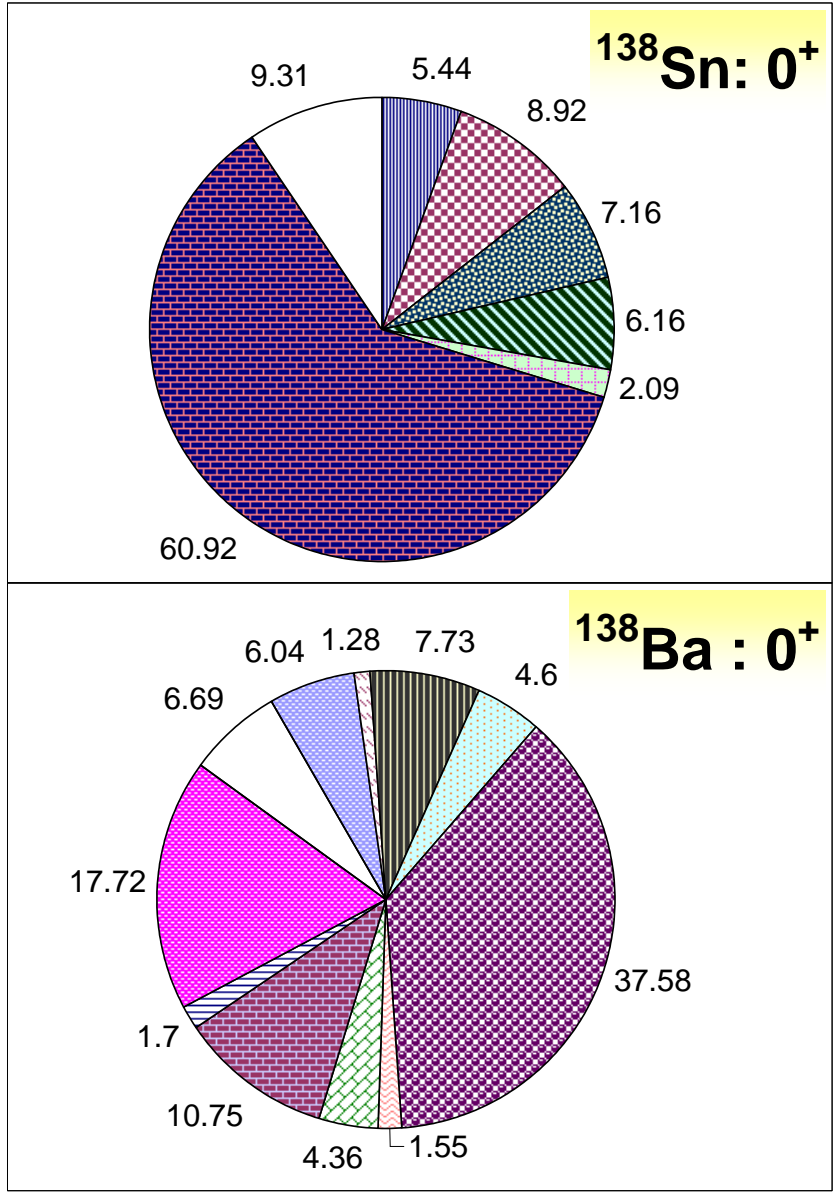}
\vspace{8.5cm}
\caption{\label{fig:fig13} (Color online) The $\pi$-chart plot of
wavefunctions of the 0$^+_1$ states of $^{138}Sn$ and $^{138}Ba$.
The  results  are  using  the  SMPN interaction. Numerical values
associated with the  different  zones  are  the  percentage  (\%)
involvement of different particle - partitions composing the wave
function.   The   zone   marked  white  indicates  the  total
contribution from all the partitions  each  of  which  contribute
less than 1\% in the wave function. }

\end{figure}
\begin{figure}[h]
\vspace{12.5cm}
\includegraphics{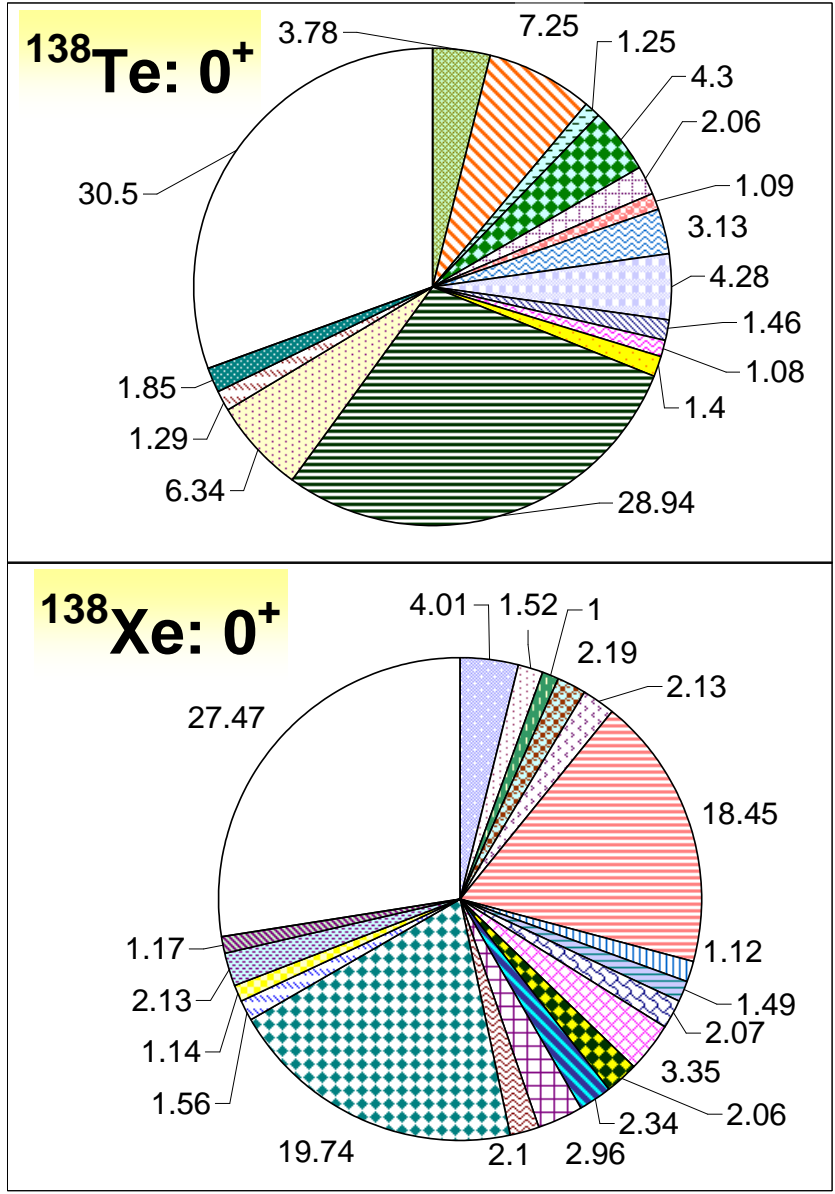}
\vspace{8.5cm}
\caption{\label{fig:fig14}(Color  online) The $\pi$-chart plot of
wavefunctions of the 0$^+_1$ states of $^{138}Te$ and $^{138}Xe$.
The results are using the SMPN interaction. See also the  caption
of Fig.13.}

\end{figure}
\end{document}